\documentstyle[graphicx,macros]{mn}
\begin{document}
%

\title[Dissecting Subhalo Evolution]
      {Dissecting the Evolution of Dark Matter Subhaloes in the Bolshoi Simulation}

\author[Frank C. van den Bosch]
       {Frank C. van den Bosch\thanks{E-mail: frank.vandenbosch@yale.edu}\\ 
        Department of Astronomy, Yale University, PO. Box 208101,
            New Haven, CT 06520-8101}


\date{}

\pagerange{\pageref{firstpage}--\pageref{lastpage}} \pubyear{2013}

\maketitle

\label{firstpage}


\begin{abstract}
  We present a comprehensive analysis of the evolution of dark matter
  subhaloes in the cosmological Bolshoi simulation. We identify a
  complete set of 12 unique evolution channels by which subhaloes
  evolve in between simulation outputs, and study their relative
  importance and demographics. We show that instantaneous masses and
  maximum circular velocities of individual subhaloes are extremely
  noisy, despite the use of a sophisticated, phase-space-based halo
  finder. We also show that subhaloes experience frequent penetrating
  encounters with other subhaloes (on average about one per dynamical
  time), and that subhaloes whose apo-center lies outside the virial
  radius of their host (the `ejected’ or `backsplash’ haloes)
  experience tidal forces that modify their orbits. This results in an
  average fractional subhalo exchange rate among host haloes of $\sim
  0.01$ Gyr$^{-1}$ (at the present time). In addition, we show that
  there are three distinct disruption channels; one in which subhaloes
  drop below the mass resolution limit of the simulation, one in which
  subhaloes `merge' with their host halo largely driven by dynamical
  friction, and one in which subhaloes abruptly disintegrate. We
  estimate that roughly 80 percent of all subhalo disruption in the
  Bolshoi simulation is numerical, rather than physical. This
  `over-merging’ is a serious road-block for the use of numerical
  simulations to interpret small scale clustering, or for any
  other study that is sensitive to the detailed demographics of
  dark matter substructure.
\end{abstract} 




\section{Introduction} 
\label{sec:intro}

During the hierarchical assembly of dark matter haloes, the inner
regions of early virialized objects often survive accretion on to a
larger system, thus giving rise to a population of subhaloes. 
Dark matter substructure plays an important role in many areas of
astrophysics. It causes time-delays (e.g. Keeton \& Moustakas 2009)
and flux-ratio anomalies (Metcalf \& Madau 2001; Brada\v{c} \etal
2002; Dalal \& Kochanek 2002) in gravitational lensing, it boosts the
dark matter annihilation signal from dark matter haloes (e.g. Diemand,
Kuhlen \& Madau 2007; Giocoli, Pieri \& Tormen 2008; Pieri, Bertone \&
Branchini 2008), and impacts the dynamics of tidal streams and
galactic discs (e.g., T\'oth \& Ostriker 1992; Taylor \& Babul 2001;
Ibata \etal 2002; Carlberg 2009). In addition, dark matter subhaloes
host satellite galaxies and the demographics of dark matter
substructure is therefore directly related to the (small scale)
clustering of galaxies. This latter idea underlies the popular
technique of subhalo abundance matching (e.g., Vale \& Ostriker 2004;
Conroy, Wechsler \& Kravtsov 2006, 2007; Guo \etal 2010; Hearin \etal
2013), which has become a prime tool for interpreting galaxy
clustering, galaxy-galaxy lensing, group multiplicity functions, and
for constraining cosmological parameters (e.g., Conroy \etal 2006;
Marin \etal 2008; Trujillo-Gomez \etal 2011; Hearin \etal 2013, 2014,
2016; Reddick \etal 2013, 2014; Zentner \etal 2014, 2016; Lehmann
\etal 2015).

Dark matter subhaloes are subjected to forces that try to dissolve it:
dynamical friction, tides from the host halo, and impulsive encounters
with other substructure. Because of the complex interplay of these
numerous, non-linear processes the formation and evolution of dark
matter substructure is best studied using numerical $N$-body
simulations. Nowadays, large cosmological simulations routinely
resolve an entire hierarchy of substructure, with haloes hosting
subhaloes, which themselves host sub-subhaloes, etc. These simulations
are used as the bedrock for semi-analytical models of galaxy formation
(e.g., Kauffmann, Nusser \& Steinmetz 1997; Croton \etal 2006; De
Lucia \& Blaizot 2007; Kang \& van den Bosch 2008; Fontanot \etal
2012), to interpret large scale structure data with the help of halo
occupation models (e.g., Conroy \etal 2006; Mar\'in \etal 2008;
Zentner \etal 2016), to study the impact of substructure on
gravitational lensing (e.g., Brada\v{c} \etal 2004; Amara \etal 2006;
Macc\'io \etal 2006; Xu \etal 2015), and even to constrain
cosmological parameters (e.g., Reddick \etal 2014).

All these methods are ultimately limited by the accuracy of the
numerical simulations used. It wasn't until the end of the 1990's that
numerical simulations started to reach sufficient mass and force
resolution to resolve a surviving population of subhaloes (Moore, Katz
\& Lake 1996; Tormen, Bouchet \& White 1997; Brainerd, Goldberg \&
Villumsen 1998; Moore \etal 1998; Ghigna \etal 1998; Tormen, Diaferio
\& Syer 1998; Klypin \etal 1999). And even today, the limiting mass
and force resolution of numerical simulations implies an over-merging
of substructure, especially near the centers of their host haloes. In
addition, the actual detection and characterization of substructure in
simulations is another important source of error.  Different subhalo
finders can yield very different results, even when applied to the
same simulation (e.g., van den Bosch \& Jiang 2016).  Hence, it is
prudent that we continue to scrutinize the numerical simulations that
we use to interpret our ever increasing amount of astrophysical data.

In this paper we examine the evolution of dark matter substructure in
a pure dark matter-only simulation. We identify a complete set of 12
unique evolution channels, by which subhaloes evolve in between
different simulation outputs, and we study the frequencies and
demographics of these evolution channels in detail. We show that
instantaneous masses and maximum circular velocities of individual
subhaloes are extremely noisy, even when these have been obtained
using sophisticated, phase-space-based subhalo finders. We also show
that subhaloes experience frequent penetrating encounters with other
subhaloes (on average about one per dynamical time), and that most
subhaloes that are on orbits that take them outside of their host (the
so-called ejected or backsplash haloes) experience tidal forces that
modify their orbital parameters. In addition, we show that there are
three distinct disruption channels; one in which subhaloes drop
below the mass resolution limit of the simulation, one in which
subhaloes `merge' with their host halo largely driven by dynamical
friction, and one in which subhaloes seem to spontaneously
disintegrate.

This paper is organized as follows. Section~\ref{sec:Method} describes
the numerical simulations and merger trees used in this study. In
\S\ref{sec:secs} and \S\ref{sec:prop} we introduce a complete set of
12 unique subhalo evolution channels, discuss their demographics and
relative importance for describing the evolution of subhaloes in the
Bolshoi simulation. Section~\ref{sec:evolution} takes a closer look at
the tidal evolution of dark matter subhaloes, focusing on tidal
stripping, penetrating encounters among subhaloes, and subhalo
disruption. We summarize our findings in \S\ref{sec:summary}.


\section{Methodology}
\label{sec:Method}

\subsection{Numerical Simulation}
\label{sec:SIM}

To study the evolution of dark matter substructure, we use the
large `Bolshoi' simulation (Klypin \etal 2011), which follows
the evolution of $2048^3$ dark matter particles using the Adaptive
Refinement Tree (ART) code (Kravtsov, Klypin \& Khokhlov 1997) in a
flat $\Lambda$CDM model with parameters $\Omega_{\rm m,0} = 1 -
\Omega_{\Lambda,0} = 0.27$, $\Omega_{\rm b,0} = 0.0469$, $h = H_0/(100
\kmsmpc) = 0.7$, $\sigma_8 = 0.82$ and $n_\rms = 0.95$ (hereafter
`Bolshoi cosmology').  The box size of the Bolshoi simulation is
$L_{\rm box} = 250 h^{-1} \Mpc$, resulting in a particle mass of
$m_\rmp = 1.35 \times 10^8 \Msunh$.

We use the publicly available halo
catalogs\footnote{http://hipacc.ucsc.edu/Bolshoi/MergerTrees.html}
obtained using the phase-space halo finder \Rockstar (Behroozi \etal
2013a), which uses adaptive, hierarchical refinement of
friends-of-friends groups in six phase-space dimensions and one time
dimension. As demonstrated in Knebe \etal (2011, 2013), this results
in a very robust identification of (sub)haloes (see also van den Bosch
\& Jiang 2016). \Rockstar haloes are defined as spheres with an
average density equal to $\Delta_{\rm vir}(z) \rho_{\rm
  crit}(z)$. Here $\rho_{\rm crit}(z) = 3 H^2(z)/8 \pi G$ is the
critical density for closure at redshift $z$, and $\Delta_{\rm
  vir}(z)$ is given by the fitting function of Bryan \& Norman (1998).
\begin{figure*}
\includegraphics[width=0.78\hdsize,angle=270]{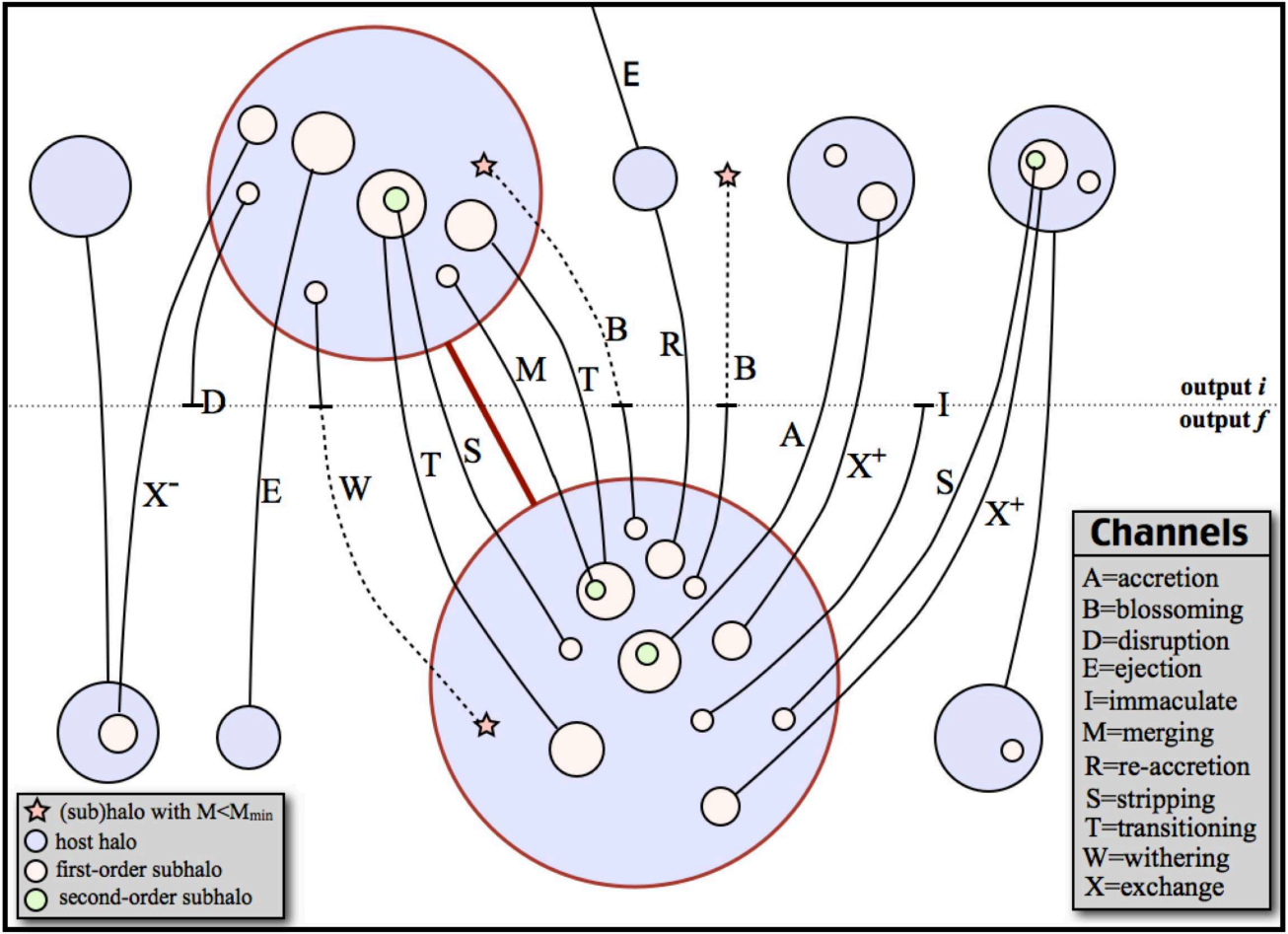}
\centering
\caption{Illustration of the various subhalo evolution channels
  (SECs). The large circles with red rims, linked by a thick red line,
  indicate host haloes in $\calS_\rmi$ (upper) and $\calS_\rmf$. All other
  sets of circles linked together correspond to a particular SEC, as
  labeled. Detailed descriptions of the various channels can be found
  in \S\ref{sec:secs} and in Table~1.}
\label{fig:secs}
\end{figure*}

We use the 19 simulation outputs at $z \leq 0.0605$, which are output
every $\Delta a = 0.003$ ($a$ is the scale factor), corresponding to
roughly $\Delta t = 42 \Myr$.  Throughout we split the halo population
in three categories: 
\begin{itemize}
\item host haloes; these are distinct haloes that are not, and never
  have been, located within the virial radius of another, more massive
  halo.
\item subhaloes; these are haloes that are located within the
  virial radius of another, more massive halo.
\item ejected haloes; these are distinct haloes whose main progenitor
  has at one or more occasions passed through the virial region of a
  more massive halo. Ejected haloes are also sometimes called
  `backsplash' haloes.
\end{itemize}
As shown in van den Bosch \etal (2014), host haloes clearly dominate,
with a fraction that increases from $\sim 70$ percent for haloes with
$\Mhalo = 10^{11} \Msunh$ to 100 percent at the massive end.  The
remainder is split roughly equally between subhaloes and ejected
haloes. Following Jiang \& van den Bosch (2016a), we also distinguish
between subhaloes of different order: we refer to subhaloes as
first-order subhaloes, to sub-subhaloes (i.e., subhaloes located
within the virial radius of another subhalo) as second-order
subhaloes, etc. An $n^{\rm th}$-order (sub)halo that hosts an
$(n+1)^{\rm th}$-order subhalo is called a {\it parent} halo of the
$(n+1)^{\rm th}$-order subhalo. Note that the masses of (sub)haloes
are defined such that the mass of an $n^{\rm th}$-order parent halo
{\it includes} the masses of its subhaloes of order $n+1$.

Throughout this paper we restrict ourselves to first-order subhaloes
with mass $m \geq \mmin \equiv 10^{9.83} \Msunh$ (corresponding to
$\geq 50$ particles per halo) that reside in host haloes with mass in
the range $10^{11.5} \Msunh \leq \Mhalo \leq 10^{15} \Msunh$, and we
discard any subhalo whose orbital energy is positive, indicating that
the subhalo is not bound to its host halo (see \S\ref{sec:prop} for
details on how the orbital energy is calculated). The reason for only
considering subhaloes with at least 50 particles is that below this
mass limit the subhalo mass functions are starting to become
incomplete due to limiting mass and/or force resolution (see van den
Bosch \& Jiang 2016).

\subsection{Merger Trees}
\label{sec:trees}

The halo catalogs at different simulation outputs, constructed using
the \Rockstar halo finder, are linked across outputs using the {\tt
  Consistent} merger tree algorithm of Behroozi \etal (2013b). We refer
the interested reader to that paper for details, but here highlight a few
aspects that are of particular relevance to this paper.

The first step of constructing the merger trees is assigning each halo
at time step $t_n$ a descendant halo at the later time step
$t_{n+1}$. This is done by identifying the halo at $t_{n+1}$ that
receives the largest fraction of the particles (excluding
substructure) of the halo in question at $t_n$. Next, the resulting
links are improved upon by using orbit integration to predict the
positions and velocities of (sub)haloes detected across adjacent
outputs.  In particular, particle-based links between halos that are
too far apart in position, velocity or mass are cut and reconnected to
more likely candidates.

Some (sub)haloes identified at time step $t_{n+1}$ may end up having
no progenitor halo at $t_n$. This may indicate that the progenitor
halo has a mass below the mass-completeness limit of the simulation,
or that the progenitor halo is missing from the halo catalog at time
step $t_n$. To account for the latter case, the merger tree algorithm
creates a place-holder halo, called a {\it phantom halo}, in the halo
catalog corresponding to $t_n$, with a mass, position and velocity
computed based on those at $t_{n+1}$ (see Behroozi \etal 2013b for
details).  Phantom haloes may be created for up to four successive time
steps to allow for cases in which the halo finder looses track of a
(sub)halo for multiple time steps. They therefore `repair' failures of
the halo finder by patching and interpolating across multiple outputs.
If a (sub)halo at $t_{n+1}$ has no progenitors in any of the four
previous time steps, it is assumed that the halo formed at $t_{n+1}$,
and the corresponding place-holder haloes in the previous four
time steps are removed from the catalogs.

If a (sub)halo at output $t_n$ has no descendant at $t_{n+1}$, or is
not the most massive progenitor of its descendant, either the halo is
a spurious `fluctuation', or it was completely disrupted during the
time interval between the two time steps (Behroozi \etal 2013b refer
to the latter as a `tidal merger'). In order to discriminate between
these two options, only those haloes that at time step $t_n$
experience a tidal acceleration $|\calT| \geq G M(r)/r^3 > 0.4 \kms
\Myr^{-1}$ comoving Mpc$^{-1}$ are considered as `true' haloes that
are disrupted. A halo that does not exceed this tidal limit is
considered a `merger fluctuation' and is removed from the halo
catalog, {\it together with all its progenitors at all previous
  timesteps}. We emphasize that these merger fluctuations are not
necessarily haloes close to the mass-completeness limit of the
simulation. For example, around $z \sim 0$ roughly $3 \times 10^{-4}$
of haloes with mass $\sim 10^{12} \Msunh$ are identified as merger
fluctuations (P. Behroozi, priv. communication).
\begin{table*}\label{tab:properties}
\caption{Subhalo Evolution Channels}
\begin{center}
\begin{tabular}{llcl}
\hline\hline
(1) & (2) & (3) & (4) \\
\hline
 T & Transition    & fwd & subhaloes in $\calS_\rmi$ that end up as subhaloes in $\calS_\rmf$ \\
 A & Accretion     & bwd & subhaloes in $\calS_\rmf$ that were host haloes at time $t_\rmi$ \\
 R & Re-accretion  & bwd & subhaloes in $\calS_\rmf$ that were ejected haloes at time $t_\rmi$ \\
 E & Ejection      & fwd & subhaloes in $\calS_\rmi$ that are host haloes at $t_\rmf$\\
 D & Disruption    & fwd & subhaloes in $\calS_\rmi$ that are disrupted at $t_\rmf$\\
 S & Stripping     & bwd & subhaloes in $\calS_\rmi$ that were subhaloes of order $n \geq 2$ at $t_\rmi$ \\
 M & Merging       & fwd & subhaloes in $\calS_\rmi$ that end up as subhaloes of order $n \geq 2$ at $t_\rmf$ \\
 W & Withering     & fwd & subhaloes in $\calS_\rmi$ that end up as subhaloes with $m < \mmin$ at $t_\rmf$ \\
 B & Blossoming    & bwd & subhaloes in $\calS_\rmf$ whose progenitor has a mass $m < \mmin$ at $t_\rmi$ \\
 I & Immaculate    & bwd & subhaloes in $\calS_\rmf$ without progenitor at $t_\rmi$ \\
 X$^{+}$ & eXchange & bwd & subhaloes in $\calS_\rmf$ that were subhaloes at $t_\rmi$ but not in the main progenitor of their current host\\
 X$^{-}$ & eXchange & fwd & subhaloes in $\calS_\rmi$ that end up as subhaloes at $t_\rmf$ but not in the descendent of their current host\\
\hline\hline
\end{tabular}
\end{center}
\medskip
\begin{minipage}{\hdsize}
  Definition of the various SEC channels that describe the evolution
  of subhaloes in set $\calS_\rmi$ at time $t_\rmi$ to that of subhaloes in
  set $\calS_\rmf$ at a later time $t_\rmf$. Here $\calS_\rmi$ and $\calS_\rmf$
  are linked by the fact that the host haloes of $\calH_\rmi$ are the
  main progenitors of the host haloes of $\calH_\rmf$. Column (1)
  indicates the letter symbol we use throughout, column (2) lists the
  name of the SEC, column (3) indicates whether it is a forward (fwd)
  or backward (bwd) channel, and column (4) gives a concise
  description.
\end{minipage}
\end{table*}

\section{Subhalo Evolution Channels}
\label{sec:secs}

Consider two simulation outputs, conveniently labeled $i$ (for
initial) and $f$ (for final), with corresponding redshifts $z_\rmf <
z_\rmi$. We start by identifying all host haloes in output $f$ in a
given mass range.  We denote this set by $\calH_\rmf$. Next we
identify the set $\calS_\rmf$ of all {\it first-order} subhaloes with
mass $m \geq \mmin$ belonging to host haloes in
$\calH_\rmf$. Subsequently we move to output $i$, and first identify
the set $\calH_\rmi$ of the main progenitors of $\calH_\rmf$. Finally,
we identify the set $\calS_\rmi$ of all {\it first-order} subhaloes
with $m\geq \mmin$ belonging to host haloes in $\calH_\rmi$. In what
follows, we use $\calS$ to denote the union of $\calS_\rmi$ and
$\calS_\rmf$ (i.e., $\calS = \calS_\rmi \cup \calS_\rmf$). Similarly,
$\calH = \calH_\rmi \cup \calH_\rmf$.
\begin{figure*}
\includegraphics[width=\hdsize]{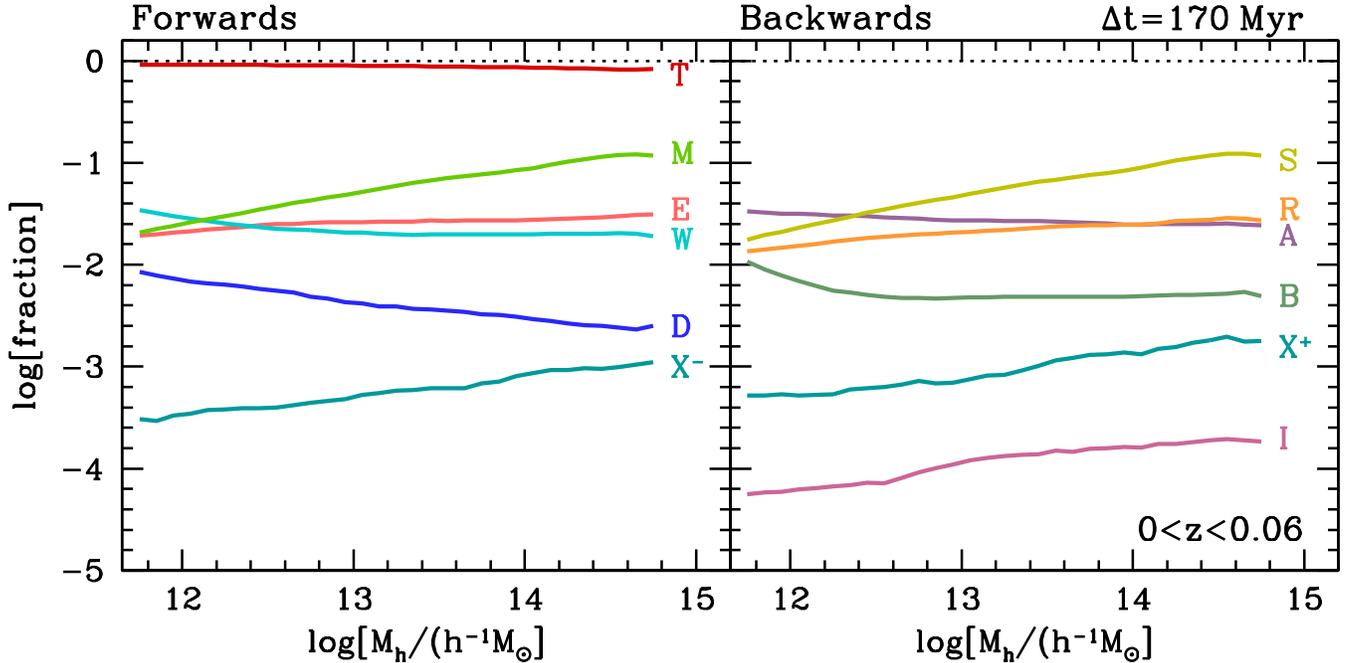}
\centering
\caption{The fractions of subhaloes that evolve in a time step of
  $\Delta t = 170\,$Myr along different SECs. Results are shown as
  function of halo mass and averaged over all simulation outputs with
  $z< 0.06$. See Fig.~\ref{fig:secs} and Table~1 for an illustration
  and description of the various SECs.}
\label{fig:fraction_mass}
\end{figure*}

The goal of this paper is to study what happens between redshifts
$z_\rmi$ and $z_\rmf$ to the subhaloes in $\calS_\rmi$ (we call this the
`forward evolution channels'), and how the subhaloes in $\calS_\rmf$ end
up in their host haloes (we call this the `backward evolution
channels'). By carefully studying the subhaloes in $\calS$, as well as
their progenitors and/or descendants, we have identified a total of 12
subhalo evolution channels (hereafter SECs); 6 forward channels and 6
backward channels (see Fig.~\ref{fig:secs} for an illustration).  This
set is both complete and unique, in that each and every subhalo in
$\calS$ is associated with one, and only one, channel. In order of
diminishing importance, these are:
\begin{itemize}
\item The Transition channel T: these are subhaloes in $\calS_\rmi$
  that end up as subhaloes in $\calS_\rmf$. As we will see, this is the
  most common evolution channel, by far. Although we refer to the
  transition channel as a forward channel, it is equally valid
  to consider it a backward channel.
\item The Accretion channel A: these are subhaloes in $\calS_\rmf$ that
  were host haloes at time $t_\rmi$. Hence, they were accreted, for the
  first time in their existence, into a host halo in the time interval
  between $t_\rmi$ and $t_\rmf$.
\item the Ejection channel E: these are subhaloes in $\calS_\rmi$ that
  are host haloes at $t_\rmf$. This channel is the inverse of the
  accretion channel. Note that ejection does not imply that the subhalo is
  no longer bound to its host halo; it merely implies that its
  center is no longer located within the host halo's virial extent.
\item the Re-accretion channel R: these are subhaloes in $\calS_\rmf$
  that were ejected haloes at time $t_\rmi$. Note that they do not
  necessarily have to have been ejected from the same host halo into
  which they are now being re-accreted.
\item the Stripping channel S: these are (first-order) subhaloes in
  $\calS_\rmf$ that were subhaloes of order $n \geq 2$ at $t_\rmi$. Recall
  that a subhalo is said to be of order $n$ if it is located within
  the extent of another subhalo of order $n-1$.
\item the Merging channel M: these are (first-order) subhaloes in
  $\calS_\rmi$ that end up as subhaloes of order $n \geq 2$ at $t_\rmf$.
  This channel is the inverse of the stripping channel.
\item the Withering channel W: these are subhaloes in $\calS_\rmi$ that
  end up as subhaloes with $m < \mmin$ at $t_\rmf$. Hence, they are
  similar to T-subhaloes, except that mass loss causes them to drop
  out of our mass-limited sample.
\item the Blossoming channel B: these are subhaloes in $\calS_\rmf$ whose
  progenitor has a mass $m < \mmin$ at $t_\rmi$.  This channel is the
  inverse of the withering channel, and refers to subhaloes that have
  increased their mass between $t_\rmi$ and $t_\rmf$ from below to above the
  mass limit of our sample.
\item the Disruption channel D: these are subhaloes in $\calS_\rmi$ that
  are disrupted at $t_\rmf$. Note that this is different from the
  withering channel, in that W subhaloes {\it do} have a descendant in
  the halo catalogue, but with a mass below the sample limit.  D
  subhaloes are identified in the \Rockstar halo catalog at $t_\rmi$ as
  having {\tt mmp}$=0$.
\item the Immaculate channel I: these are subhaloes in $\calS_\rmf$
  without progenitor at $t_\rmi$. This channel is the inverse of the
  disruption channel, and identifies subhaloes that appear in the
  \Rockstar halo catalog at $t_\rmf$ as a subhalo without having an
  (identified) progenitor at $t_\rmi$. I subhaloes are identified in the 
  halo catalog at $t_\rmf$ as having {\tt num}$_{\rm prog}<1$ and {\tt PID}$>0$.
\item the positive eXchange channel X$^{+}$: these are subhaloes in
  $\calS_\rmf$ that were subhaloes at $t_\rmi$ but not in the main
  progenitor of their current (at $t_\rmf$) host halo.
\item the negative eXchange channel X$^{-}$: these are subhaloes in
  $\calS_\rmi$ that end up as subhaloes at $t_\rmf$ but not in the
  descendent of their current (at $t_\rmi$) host halo. This channel is
  the inverse of the X$^{+}$ channel.
\end{itemize}
A list of concise descriptions for each SEC can be found in Table~1. 
In what follows we will refer to each channel with the capital letter
indicated above and in the first column of Table~1. Finally, it is
important to be aware of the following: Ejection and re-accretion can
go unnoticed if both happen within the time-interval between the two
simulation outputs. In that case the subhalo is assigned to the
transition channel, T. In addition, if a subhalo is ejected and
subsequently captured by another host halo, the subhalo is assigned to
the negative exchange channel, X$^{-}$, rather than the ejection
channel.

\subsection{The Relative Importance of different SECs}
\label{sec:fractions}

Given two simulation outputs, at $t_\rmi$ and $t_\rmf$, a range in host halo
mass, and the corresponding sets of subhaloes, $\calS_\rmi$ and
$\calS_\rmf$, we determine for each subhalo in $\calS$ to which of the
twelve SEC channels it belongs. Next we compute the fractional
contributions (by number) for each of these channels, which are
defined as
\begin{equation}
f_\rmc \equiv \left\{ 
\begin{array}{ll}
N_\rmc/N_\rmi & \mbox{if forward channel} \\
N_\rmc/N_\rmf & \mbox{if backward channel}
\end{array} \right.
\end{equation}
Here $N_\rmc$ is the number of subhaloes belonging to channel `c'
and $N_\rmi$ and $N_\rmf$ are the total number of subhaloes in $\calS_\rmi$
and $\calS_\rmf$, respectively.

Fig.~\ref{fig:fraction_mass} plots $f_\rmc$ for the various forward
(left-hand panel) and backward (right-hand panel) channels as function
of the mass of the host halo at the later output (i.e., at
$t_\rmf$). These fractions correspond to a time interval in between
simulation outputs of $\Delta t = 170\Myr$.  Among the 19 simulation
outputs used here, there are a total of 14 pairs that are (roughly)
separated by this time-interval, and the fractions shown are the
averages among those 14 pairs. Note that the time intervals between
$t_\rmi$ and $t_\rmf$ considered throughout this paper are sufficiently
short that the host halo masses evolve very little. For instance, for
$\Delta t = 170\Myr$ the average host halo in our sample only
increases its mass by 0.9 percent.

By far the dominant channel is the transition channel T, which
describes subhaloes that simply continue to orbit within one and the
same host halo. Next up in order of importance, are the stripping, S,
and merging, M, channels. These describe subhaloes that either
decrease or increase their order between $t_\rmi$ and $t_\rmf$,
respectively. Both contribute more in more massive host haloes,
reaching fractional contributions of $\sim 10$ percent in cluster-size
hosts. The fractional contribution of accretion of new subhaloes, A,
is similar to that of ejection, E, and re-accretion, R, which as we
will see in \S\ref{sec:AER} is in agreement with simple expectations
based on the orbits of subhaloes at infall.

Another channel that contributes roughly equally is the withering
channel, W, which describes subhaloes that experience mass loss
between $t_\rmi$ and $t_\rmf$ such that at $t_\rmf$ their mass has dropped
below the limit of 50 particles per halo. The inverse of the withering
channel is the blossoming channel, B, which describes subhaloes whose
mass has increased from below $\mmin$ to above $\mmin$ in the time
interval between $t_\rmi$ and $t_\rmf$. Blossoming subhaloes contribute
between $\sim 1\%$ in low mass host haloes and $\sim 0.5\%$ at the
massive end, and as we will see, are mainly a `by-product' of noise in
the subhalo mass assignment (see \S\ref{sec:WB}). Channel D describes
subhaloes that disrupt between $t_\rmi$ and $t_\rmf$, and contributes
between one percent ($\Mhalo \sim 10^{12} \Msunh$) and 0.3 percent
($\Mhalo \sim 10^{15}\Msunh$).
\begin{figure*}
\includegraphics[width=\hdsize]{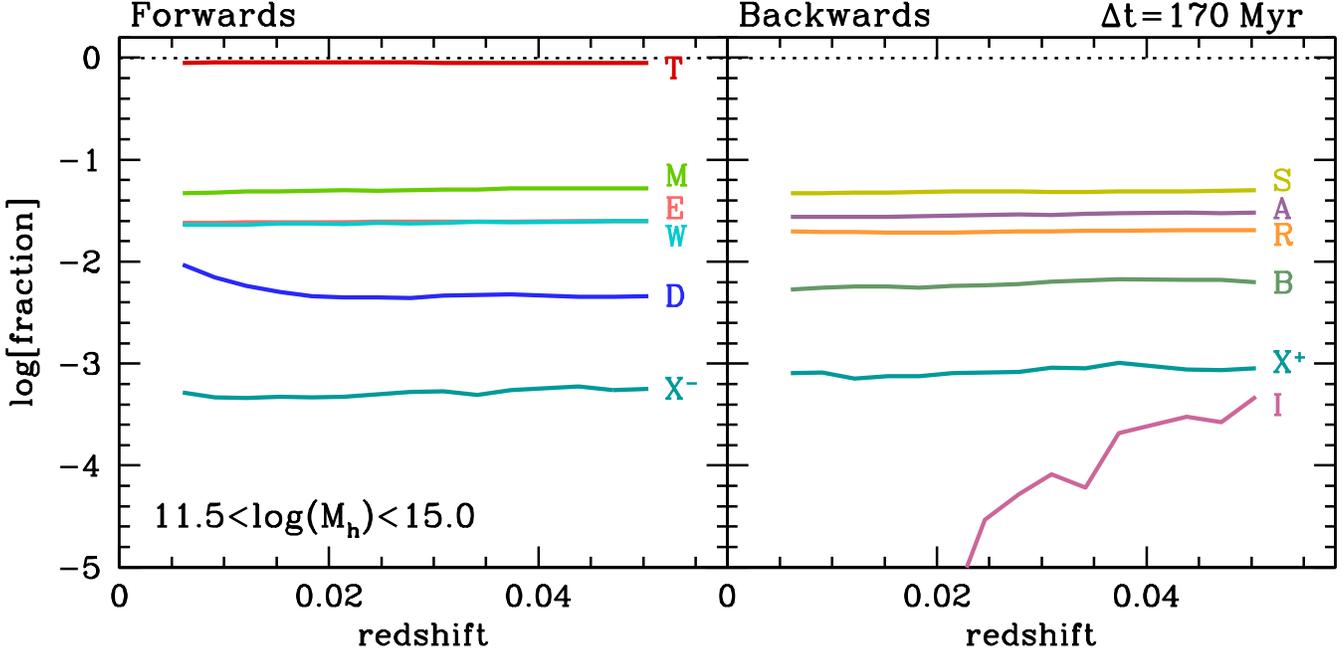}
\centering
\caption{Same as Fig.~\ref{fig:fraction_mass}, but this time the
  fractions are computed for host haloes in the range $10^{11.5} \leq
  \Mhalo \leq 10^{14.5} \Msunh$ and plotted as a function of the
  average redshift between the two epochs separated by $\Delta t =
  170\Myr$.}
\label{fig:fraction_redshift}
\end{figure*}

The positive and negative exchange channels, X$^{+}$ and X$^{-}$,
respectively, typically contribute only $\lta 0.1\%$, with a weak
dependence on host halo mass. Exchange channels basically describe
`ejection' followed by `re-accretion', except that the subhalo is
re-accreted into another host halo than the one from which it was
ejected. As is evident from Fig.~\ref{fig:secs}, another contribution
to the X$^{+}$ channel is from a combination of accretion plus
stripping.  If both processes occur within the same time-interval,
this subhalo will be assigned to the X$^{+}$ channel. Finally, the
least significant channel is that of the `immaculate' subhaloes, I,
which are subhaloes in $\calS_\rmf$ that have no progenitor halo at
$t_\rmi$. Either they formed as a host halo, and were subsequently
accreted (all within the interval $\Delta t = t_\rmf-t_\rmi$), or they
popped into existence as a subhalo. Since the latter is unphysical, it
corresponds to a fluke of the (sub)halo finder or the merger tree
algorithm. In particular, if the (sub)halo finder is unable to find
the subhalo in more than four consecutive time-steps, then the
phantom-patching (described in \S\ref{sec:trees}) fails and the
subhalo will re-appear as an immaculate subhalo.  In a time interval
$\Delta t = 170\Myr$, the fractional contribution of immaculate
subhaloes (averaged over the redshift range $0 < z < 0.06$), is only
$\sim 10^{-4}$.

Fig.~\ref{fig:fraction_redshift} shows the same $\Delta t = 170\Myr$
SEC fractions as in Fig.\ref{fig:fraction_mass}, but this time as
function of redshift. Except for the I and D channels, all SEC
fractions are independent of redshift, at least over the small
redshift range considered here ($z < 0.06$). The disruption channel is
redshift independent for $z \gta 0.018$, but then increases slightly
towards $z=0$. The reason for this behavior is discussed in detail in
\S\ref{sec:DPI} below, but briefly, it arises because at low redshift
there is not enough temporal information into the future to test
whether the subhalo has disappeared from the catalog due to issues
with the halo finder, or whether the disruption is real.  The
immaculate channel reveals a dramatic redshift dependence.  There are
zero immaculates in the nine \Rockstar halo catalogs with $z <
0.026$. However, at higher redshifts the fractional contribution of
immaculates increases to almost 0.1\% at $z=0.05$.  This redshift
dependence arises from the fact that the {\tt Consistent} merger tree
algorithm of Behroozi \etal (2013b) removes subhaloes that are tracked
for fewer than 10 time steps and never orbit outside of the virial
radius of their host haloes. This effectively removes all immaculates
from the halo catalogs in the outputs close to $z=0$.

\subsection{Dependence on Time Step}
\label{sec:timestep}

The fractions discussed above are all functions of host halo mass,
$\Mhalo$, redshift, $z$, and the time-interval, $\Delta t$, between
the simulation outputs. In the discussion above, we examined the
dependencies on $\Mhalo$ and $z$ for fixed $\Delta t \sim 170 \Myr$.
We now focus on the dependence on $\Delta t$. 

\begin{figure*}
\includegraphics[width=\hdsize]{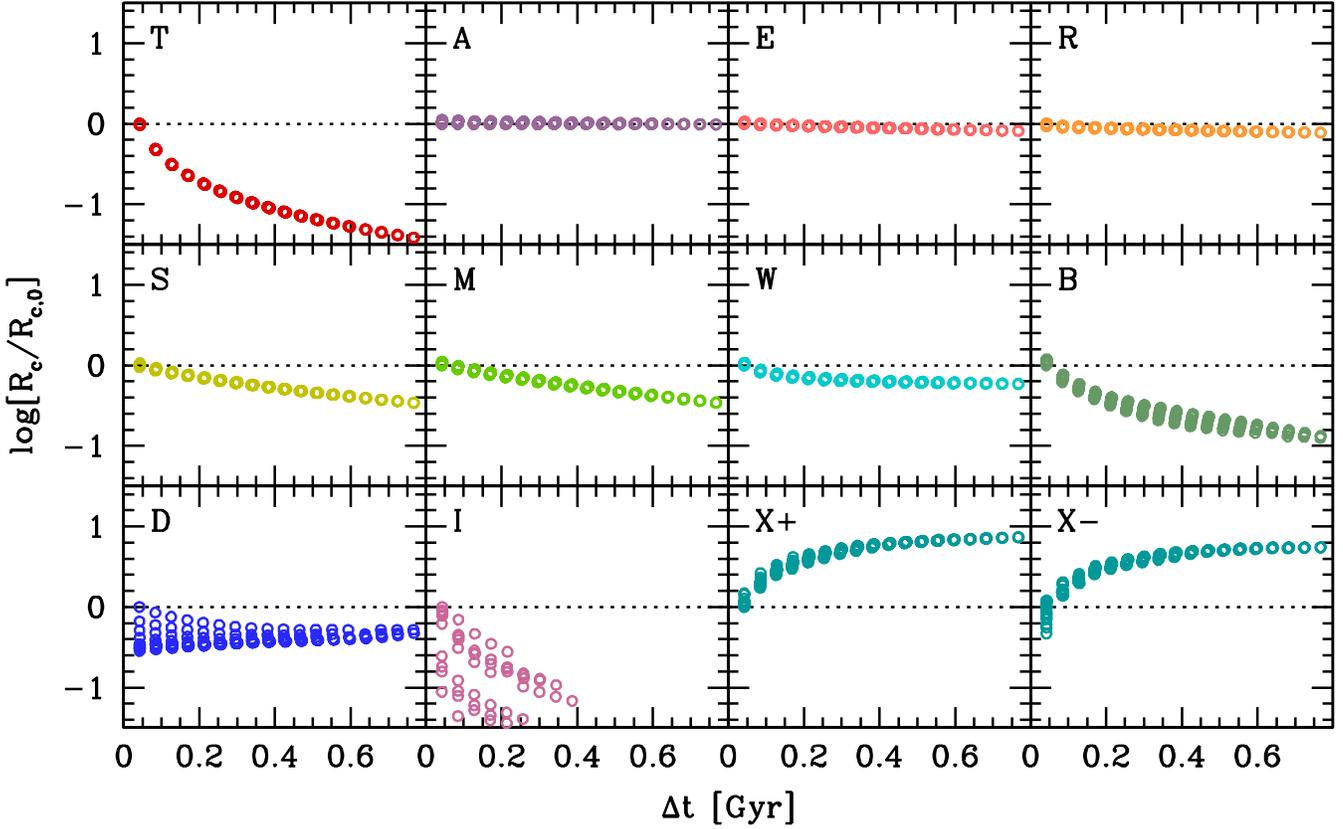}
\centering
\caption{The logarithm of the normalized SEC rate (SEC fraction
  divided by the time interval, $\Delta t$, between two outputs,
  normalized by that between the final two outputs) as function of
  $\Delta t$ (in Gyr). Each panel corresponds to a different SEC, as
  indicated in the upper, left corner, and shows the results for all
  117 output-pairs. If $R_\rmc/R_\rmc,0 = 1$ for all $\Delta t$ it
  indicates a constant rate, and thus a fractional SEC contribution
  that increases linearly with $\Delta t$. Note that we have adopted
  the same color-coding as in Figs.~\ref{fig:fraction_mass}
  and~\ref{fig:fraction_redshift}.}
\label{fig:time}
\end{figure*}

Using the 19 simulation outputs, we compute the various SEC fractions
for all 171 different output pairs. Dividing by the time interval
between the two outputs yields the fractional rate $\calR_\rmc \equiv
f_\rmc/\Delta t$, where `c' can be any of the twelve SECs. We
normalize these fractional rates by those for the last two outputs
(corresponding to $z=0.003$ and $z=0$), and plot the resulting
$\log[\calR_\rmc/\calR_{\rmc,0}]$ for each of the 171 different pairs
as function of the corresponding $\Delta t$ in Fig.~\ref{fig:time}.
Note that since the outputs are separated by almost identical time
intervals, there are typically multiple output-pairs that correspond
to the same $\Delta t$ (i.e., there are $19-n$ pairs with $\Delta t
\simeq n \times 42\Myr$, with $n=1,2,...18$).

The transition channel, T, has a fractional rate that decreases
precipitously with increasing $\Delta t$. This is simply a consequence
of the fact that
\begin{equation}
f_\rmT = 1 - \sum_{{\rm fwd} \ne \rmT} f_\rmc
\end{equation}
where the summation is over all forward channels other than T.  Since
the fractional contributions from all those channels increase with
$\Delta t$, that of T must decrease.

The normalized, fractional rates for the accretion, ejection, and
re-accretion channels are, to good approximation, independent of
$\Delta t$.  This indicates that their corresponding fractions are
proportional to $\Delta t$. The fractional rates of stripping and
merging, on the other hand, decrease with increasing $\Delta t$. As we
discuss in \S\ref{sec:SM}, this is a consequence of the fact that the
S and M channels are strongly correlated temporally, in that a subhalo
that recently experienced stripping (merging) is more likely to
experience merging (stripping). The same holds for the blossoming
and withering channels, B and W.

Both the D and I channels reveal a dispersion in
$\calR_\rmc/\calR_{\rmc,0}$ at fixed $\Delta t$. This is simply a
manifestation of the redshift dependence of their fractions (see
Fig.~\ref{fig:fraction_redshift}).  Note that $\calR_{\rmc, 0}=0$ for
the immaculates, and we have therefore normalized its fractional rates
by that for the {\it first} two outputs (corresponding to $z=0.057$
and $z=0.060$) instead.  The resulting normalized rates decrease with
increasing $\Delta t$, which arises from the fact that $f_\rmI = 0$
for $z < 0.026$ and strongly increases with redshift thereafter. In
the case of disruption, $f_\rmD$ increases slightly towards $z=0$.  If
we only include outputs at $z > 0.018$, for which $f_\rmD$ is
independent of redshift, we find only a weak increase of the
normalized, fractional disruption rate with increasing $\Delta t$.

Finally, the exchange channels X$^{+}$ and X$^{-}$ are `composite'
channels, in that they represent combinations of two events; either
ejection combined with re-accretion, or accretion combined with
stripping (cf. Fig.~\ref{fig:secs}). Given that there is a
characteristic time interval, $(\Delta t)_\rmc$, between these two
events, we expect the fractional exchange rates to increase with
$\Delta t$, until $\Delta t \gta (\Delta t)_\rmc$, after which the
fractional rate is expected to level out. The lower-right panels of
Fig.~\ref{fig:time} confirm this behavior, and suggest that $(\Delta
t)_\rmc \sim 0.6 \Gyr$. If we use the asymptotic rates for $\Delta t >
0.6 \Gyr$ as estimate of the effective rate at which subhaloes are
exchanged among host haloes, and the take account of the weak halo
mass dependence seen in Fig.~\ref{fig:fraction_mass}, we infer
fractional subhalo exchange rates of 
\begin{eqnarray}\label{rateX}
  \calR(\rmX^{+}) & = & 0.011 \, \Gyr^{-1} \left({M_\rmh \over 10^{13} \Msunh}
  \right)^{0.2} \nonumber \\
  \calR(\rmX^{-}) & = & 0.008 \, \Gyr^{-1} \left({M_\rmh \over 10^{13} \Msunh}
  \right)^{0.2}
\end{eqnarray}
Hence, over a Hubble time a few percent of subhaloes (and thus
satellite galaxies) is expected to have changed their host halo.  Note
that the positive exchange rate (host gains a subhalo from another
host) is slightly higher than the negative exchange rate (host looses
a subhalo to another host). This is a consequence of the weak mass
dependencies, and the fact that we have restricted ourselves to fairly
massive host haloes with $10^{11.5} \Msunh \leq \Mhalo \leq 10^{15}
\Msunh$.
\begin{figure*}
\includegraphics[width=\hdsize]{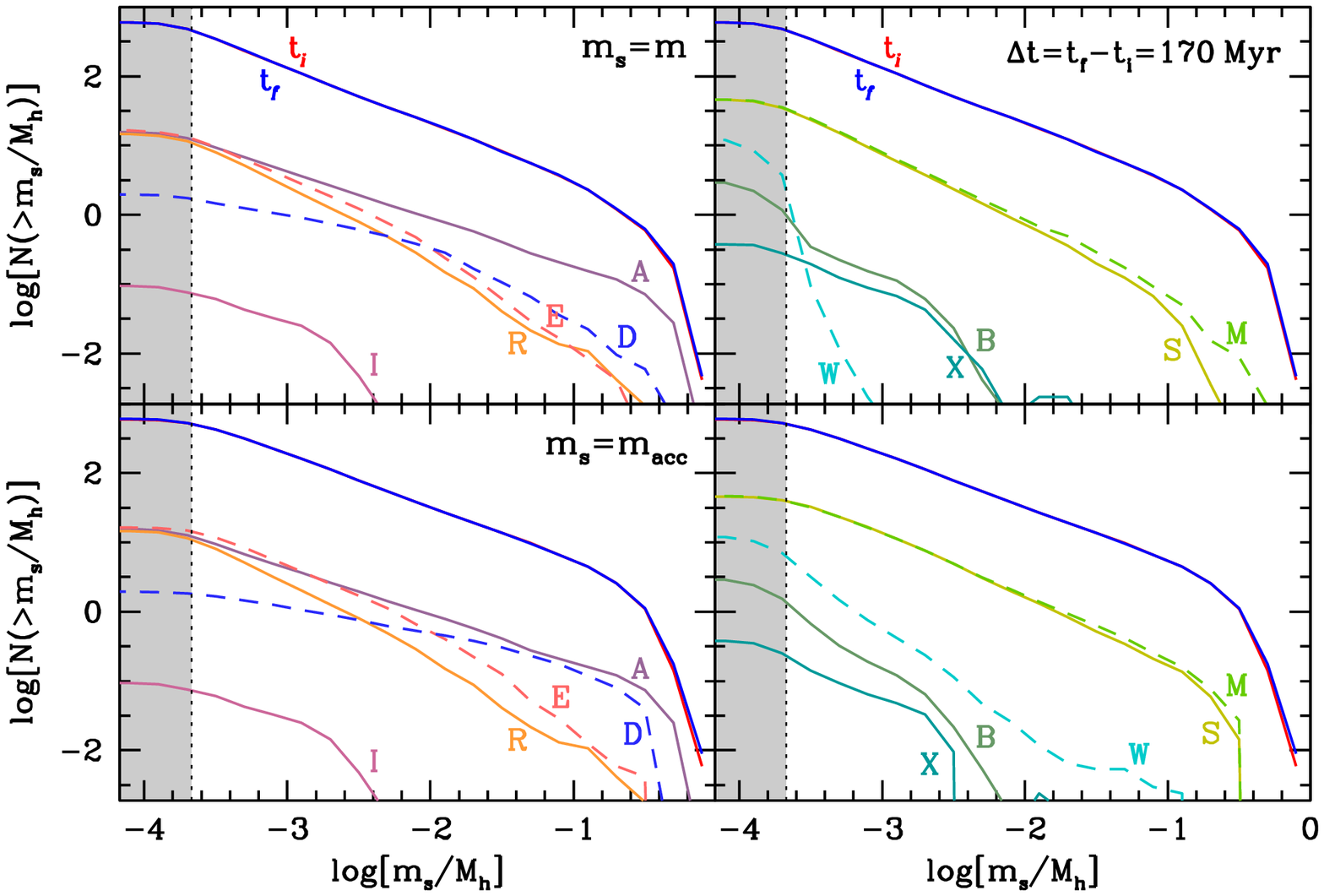}
\centering
\caption{Cumulative mass functions for the various SECs for subhaloes
  that reside in host haloes with $13.5 \leq \log[\Mhalo/(\Msunh)]
  \leq 14.0$ and for $\Delta t = t_\rmf - t_\rmi = 170 \Myr$. Upper
  and lower panels show the results for the present-day (evolved)
  subhalo mass, $m$, and the mass at accretion, $\macc$, respectively,
  both normalized to $\Mhalo$. Solid and dashed curves correspond to
  channels that add and remove subhaloes, respectively. The
  grey-shaded regions mark where $m_\rms < \mmin$. Labeling and
  color scheme is the same as in all previous figures.}
\label{fig:massfunc}
\end{figure*}

\subsection{Mass Functions}
\label{sec:massfunc}

An important diagnostic for the various SECs are their mass functions;
both in terms of their instantaneous mass, $m$, and their mass at
accretion, $\macc$. The blue and red curves in Fig.~\ref{fig:massfunc}
show the cumulative subhalo mass functions for subhaloes residing in
host haloes with $13.5 \leq \log[\Mhalo/(\Msunh)] < 14.0$. Blue and
red correspond to different epochs, $t_\rmf$ and $t_\rmi$,
respectively, that are separated by $170\Myr$. Clearly, over this time
interval there is little to no {\it net} evolution in the subhalo mass
function. The other curves indicate the mass functions of subhaloes
associated with the various SEC channels, as indicated. Solid and
dashed curves correspond to channels that add and remove subhaloes,
respectively. Note that we have split the channels over two different
panels in order to avoid clutter. The masses of the subhaloes, $m$,
are normalized by the masses of their host haloes, $M$, while the
vertical dotted lines corresponds to $m/M = \mmin/10^{14} \Msunh$,
below which the cumulative mass functions start to asymptote due to
the fact that our sample only includes subhaloes with $m \geq
\mmin$. We will discuss the mass functions of the various SEC channels
below, when we discuss the individual channels in detail.

\section{Subhalo Properties along Individual Channels}
\label{sec:prop}

Having discussed the fractional contributions of the various SECs to
the evolution of subhaloes, we now take a closer look at subhaloes
along each individual channel. In particular, we compare subhaloes
following different SECs in terms of their orbits as well as their
properties both at the present and at the epoch of accretion.

We characterize orbits using the specific orbital energy, $E = V^2/2 +
\Phi(r)$, and the orbital angular momentum. Here $r$ is the
halo-centric distance of the subhalo from the center of its host halo,
\begin{equation}\label{velocity}
V = \vert \vec{v}_{\rm host} - \vec{v}_{\rm sub} \vert
\end{equation}
is the speed of the subhalo with respect to the center of its host
halo, and $\Phi(r)$ is the potential energy at the location $r$ of the
subhalo. We compute the latter assuming that the host halo is a
spherical NFW (Navarro, Frenk \& White 1997) halo, so that
\begin{equation}\label{PhiNFW}
\Phi(r) = -V^2_{\rm vir} \, {\ln(1+cx) \over f(c) \, x}\,.
\end{equation}
Here $V_{\rm vir} = \sqrt{G M/r_{\rm vir}}$ is the circular velocity
of the host halo at its virial radius, $x=r/r_{\rm vir}$,
\begin{equation}\label{fx}
f(x) = \ln(1+x) - {x \over 1+x}\,,
\end{equation}
and $c$ is the host halo's concentration parameter. Recall that
subhaloes on unbound orbits ($E > 0$) are discarded from our sample
(see \S\ref{sec:SIM}).

The red histograms in Fig.~\ref{fig:AER} (reproduced, for comparison,
in Figs.~\ref{fig:SM}--\ref{fig:DPI}) show the distributions of a
total of six properties for {\it all} subhaloes, independent of the
SEC channel to which they belong. From top-left to bottom-right these
properties are:

\begin{itemize}

\item $r_\rmc(E)/r_{\rm vir}$, the radius of the circular orbit
  corresponding to the orbital energy $E$, expressed in terms of the
  virial radius of the host halo.

\item $\eta$, the orbital circularity, defined as the ratio of the
  orbital angular momentum, $L$, and the angular momentum $L_\rmc(E)$
  corresponding to a circular orbit of energy $E$. Radial and circular
  orbits have $\eta = 0$ and 1, respectively.

\item $r/r_{\rm vir}$, the current halo-centric distance of the
  subhalo in units of the virial radius of the host halo.

\item $z_{\rm acc}$ the redshift of accretion into the main
  progenitor of the current host halo.

\item $m/m_{\rm acc}$, the ratio of the mass of the subhalo at the
  epoch under investigation to that at accretion.

\item $K/|W|$, the virial parameter with $K$ and $W$ the total kinetic
  and potential energy of the subhalo. If the subhalo is in virial
  equilibrium with negligible surface pressure, then we expect that
  $K/|W|=1/2$. Note that $W$ is computed ignoring any external
  potential, such as that from the host halo.

\end{itemize}

\subsection{Subhaloes in Transition}
\label{sec:T}

\begin{figure*}
\includegraphics[width=\hdsize]{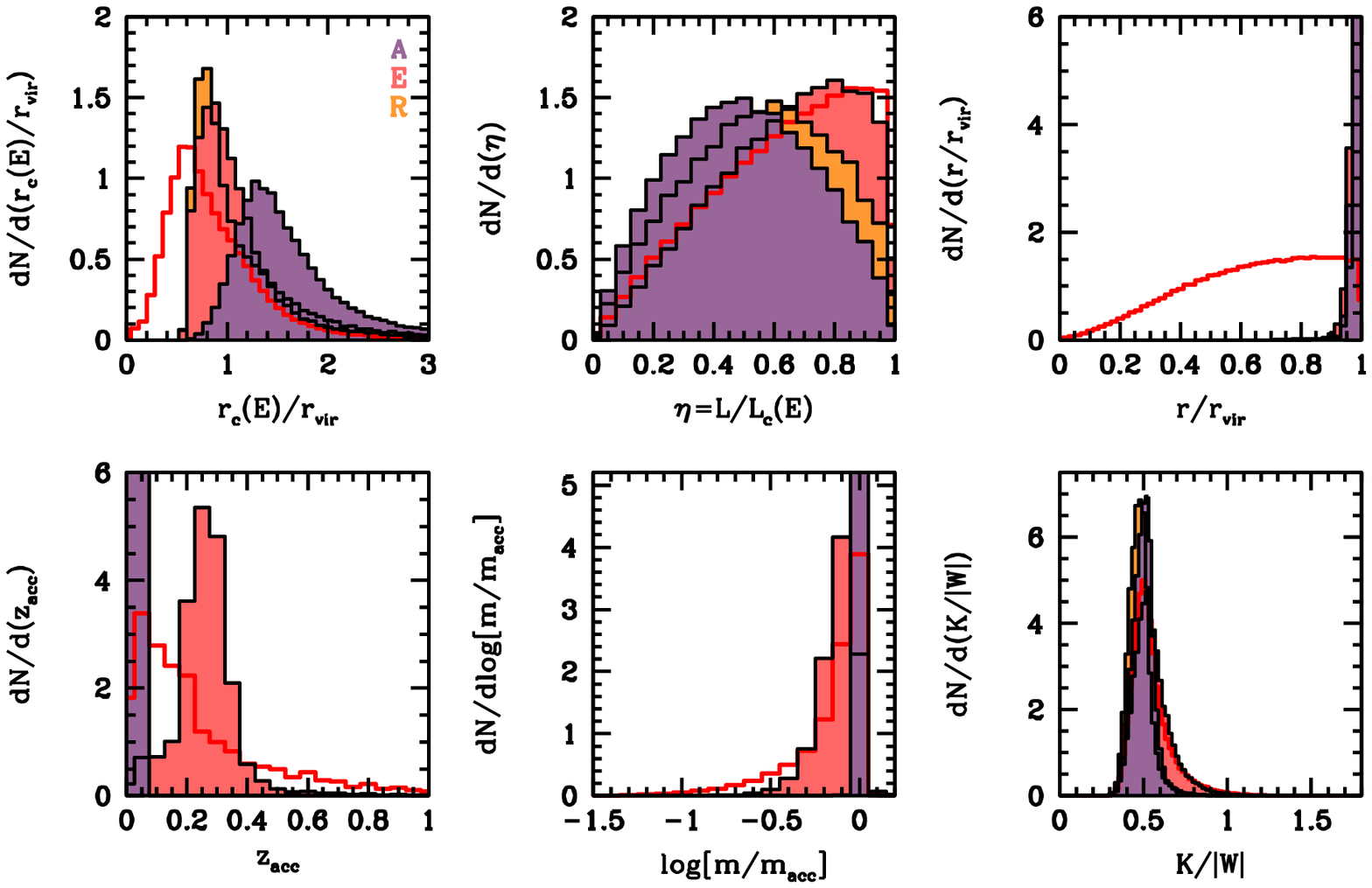}
\centering
\caption{From left to right and top to bottom, the various panels show
  normalized distributions of the orbital energy expressed as
  $r_\rmc(E)/r_{\rm vir}$, orbital circularity $\eta$, instantaneous
  radius $r/r_{\rm vir}$, accretion redshift $z_{\rm acc}$, the
  logarithm of the mass ratio $m/m_{\rm acc}$, and the virial ratio
  $K/|W|$. Shaded histograms correspond to subhaloes belonging to the
  accretion channel, A, the ejection channel, E, and the re-accretion
  channel, R, as indicated, while the red histogram corresponds to
  {\it all} subhaloes, and is indistinguishable from that for
  transition (T) subhaloes. Note that, by definition, A and R
  subhaloes have $r \simeq r_{\rm vir}$ $m \simeq m_{\rm acc}$ and
  $z_{\rm acc}$ equal to the simulation output under consideration
  (all of which have $z < 0.06$.}
\label{fig:AER}
\end{figure*}

\begin{figure*}
\includegraphics[width=\hdsize]{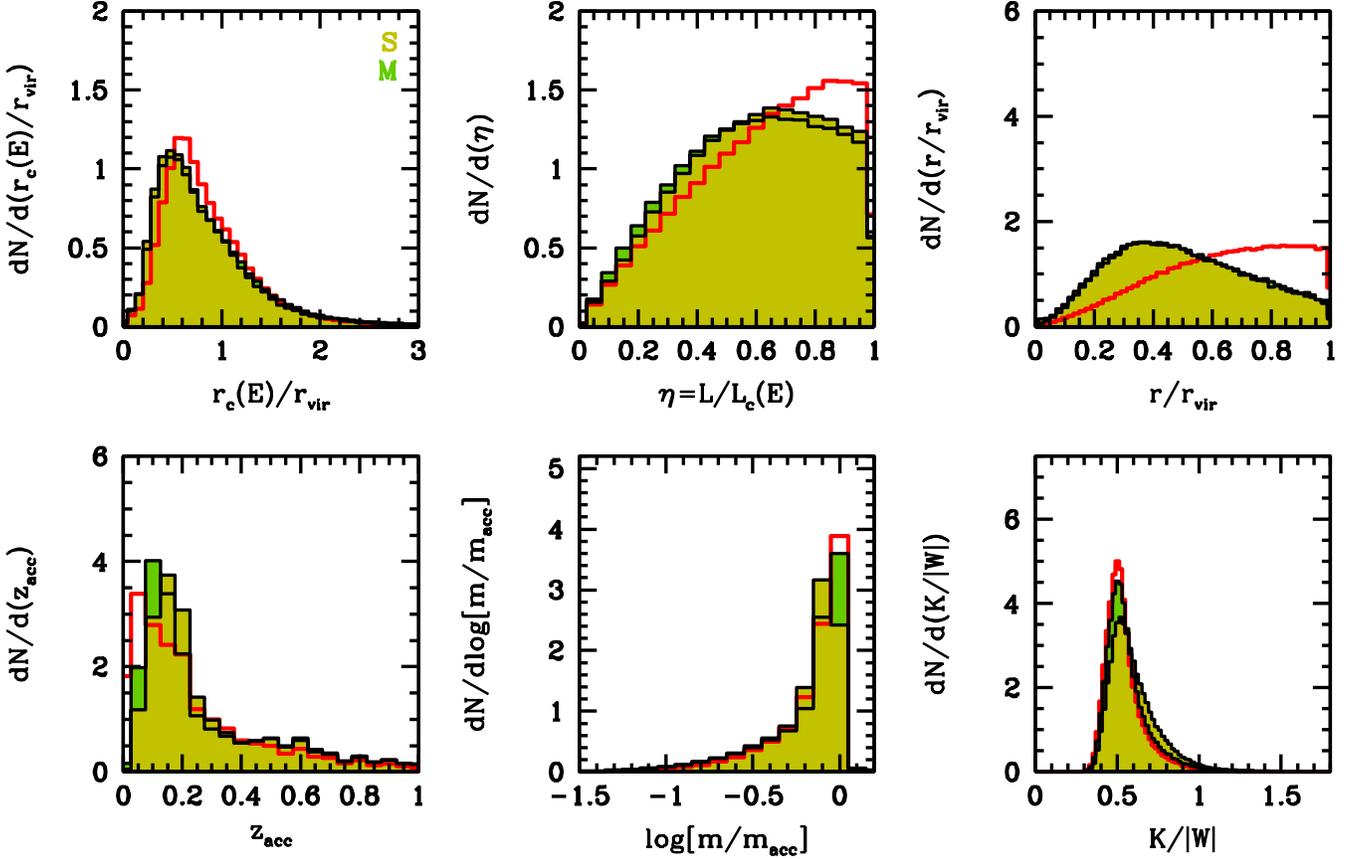}
\centering
\caption{Same as Fig.~\ref{fig:AER}, but for the stripping (S) and
  merging (M) channels. Note that S and M subhaloes are virtually
  indistinguishable. Their properties are also very similar to those
  of all subhaloes (red histogram), although their distribution of
  $r/r_{\rm vir}$ is clearly offset towards smaller halo-centric
  distances.}
\label{fig:SM}
\end{figure*}

By far the dominant SEC is the transition channel, T. For $\Delta t
\simeq 42\Myr$ more than 96 percent of the subhaloes evolve along this
channel.  Consequently, the properties of T subhaloes are virtually
indistinguishable from those of {\it all} subhaloes, and are therefore
well represented by the red histograms in
Figs.~\ref{fig:AER}--\ref{fig:DPI}. As is evident from the upper
left-hand panel in those figures, a large fraction of T subhaloes are
on orbits whose energy corresponds to a radius $r_\rmc(E)$ that is
larger than the host halo's virial radius.  Furthermore, T subhaloes
are on orbits whose circularity, $\eta$, is strongly skewed towards
more circular orbits.  For an orbit with energy $E$ and angular
momentum $L$ in a spherical potential, the peri-center, $r_\rmp$, and
apo-center, $r_\rma$, are the roots for $r$ of
\begin{equation}
{1 \over r^2} + {2 [\Phi(r) - E] \over L^2} = 0
\end{equation}
(Binney \& Tremaine 2008). Using the distributions of $E$ and $L$, we
find that 59 percent of T subhaloes have $r_\rma > r_{\rm vir}$ so
that their orbit will take them beyond the host halo's (current)
virial radius (i.e., they will at some point be `ejected').

The distribution of halo-centric distances, $r/r_{\rm vir}$, is less
centrally concentrated than what is expected for an NFW profile. This
indicates that subhaloes are spatially anti-biased with respect to the
dark matter distribution, a result that is well established (e.g.,
Ghigna \etal 1998, 2000; Gao \etal 2004; Diemand, Moore \& Stadel
2004; Springel \etal 2008; Jiang \& van den Bosch 2016b). The $z_{\rm
  acc}$ distribution is skewed towards small $z_{\rm acc}$, indicating
that most surviving subhaloes, at any given time, have been accreted
fairly recently (e.g., Zentner \& Bullock 2003; Gao \etal 2004; van
den Bosch, Tormen \& Giocoli 2005; Jiang \& van den Bosch 2016a). We
find that 50 percent of all $z=0$ subhaloes in our sample have $z_{\rm
  acc} < 0.164$.

The $m/m_{\rm acc}$ distribution of subhaloes in the Bolshoi
simulation is remarkably narrow.  As is well known, the retained mass
fraction $m/m_{\rm acc}$ is strongly correlated with $z_{\rm acc}$, in
that haloes that were accreted earlier have experienced more mass
loss, and thus have smaller $m/m_{\rm acc}$ (e.g., Gao \etal 2004;
Zentner \etal 2005; van den Bosch \etal 2005, 2016). We find that 1.3
(0.02) percent of the subhaloes in our sample have $m/m_{\rm acc} <
0.1$ ($< 0.01$). As shown in Jiang \& van den Bosch (2016a), the
absence of subhaloes with small $m/m_{\rm acc}$, and thus large
$z_{\rm acc}$, is mainly a consequence of subhalo disruption.

Finally, the distribution of the virial ratio is narrow, and centered
around $K/|W| \simeq 0.5$, corresponding to virial equilibrium. The
distribution is skewed towards higher values.  As shown in van den
Bosch \etal (2016), subhaloes experience a tidal shock near
peri-center which temporarily boosts their $K/|W|$. Since the total
binding energy of a dark matter subhalo is $E = K + W$, a virial ratio
of $K/|W|>1$ corresponds to a positive binding energy; i.e., a system
that is likely to be close to tidal disruption. Among our sample of
subhaloes, $\sim 0.9$ percent have $K/|W| > 1$.

\subsection{Accretion, Ejection and Re-accretion}
\label{sec:AER}

The accretion, ejection and re-accretion channels are closely related.
In fact, they signal different events in the life of a typical subhalo
(e.g., Lin, Jing \& Lin 2003; Gill, Knebe \& Gibson 2005; Sales \etal
2007; Ludlow \etal 2009).  Accretion corresponds to the moment a
subhalo enters the virial extent of its host halo, and if its orbital
energy and angular momentum are conserved quantities, its orbit will
once again take it outside of the host halo's virial radius
(`ejection').  Hence, every subhalo is expected to be ejected again,
unless either (i) the subhalo is disrupted during its first
peri-centric passage, (ii) the subhalo experiences significant
dynamical friction, or (iii) the host halo grows substantially during
the orbital period, so that its virial radius becomes larger than the
subhalo's apo-center.  In fact, as we now demonstrate, all three
processes play a role and have a dramatic impact on the freshly
accreted population of subhaloes.

We start our discussion by comparing subhaloes at accretion and at
ejection. As is evident from the lower-left panel of
Fig.~\ref{fig:AER}, subhaloes that are being ejected have a $z_{\rm
  acc}$ distribution that peaks around $0.25$ (see also van den Bosch
\etal 2016). Hence, at present there is on average about 3 Gyr between
accretion and ejection.

The upper-left and upper-middle panels of Fig.~\ref{fig:AER} reveal a
dramatic difference in the orbital properties of A and E subhaloes.
Subhaloes at accretion have a broad, skewed distribution of
$r_\rmc(E)/r_{\rm vir}$. The median is $1.26$ and 11.2 percent of the
A subhaloes have $r_\rmc(E)/r_{\rm vir} > 3$.  For comparison, the
median for E subhaloes is $1.0$, while only $\sim 4$ percent have
$r_\rmc(E)/r_{\rm vir} > 3$. The $\eta$ distributions are also very
different.  While that of subhaloes at accretion is remarkably
symmetric around $\eta \sim 0.5$, in good agreement with previous
studies (e.g., Tormen 1997; Zentner \etal 2005; Wetzel 2011; Jiang
\etal 2015), the distribution of orbital circularities for subhaloes
at ejection is strongly skewed towards more circular orbits.  In fact,
the $\eta$ distribution of $E$ subhaloes is very similar to that of
{\it all} subhaloes. Note that this is contrary to naive expectations:
subhaloes on more radial orbits (i.e., with smaller $\eta$), have
larger apo-centric distances than orbits of the same energy with
larger $\eta$. Hence, one might have expected that subhaloes with
lower values of $\eta$ are more likely to experience ejection.

What causes this dramatic change in the orbital properties, over a
time interval of only $\sim 3\Gyr$? As shown in Wetzel (2011), the
evolution in orbital parameters of accreting subhaloes between
$z=0.25$ and $z=0$ is negligible. Hence, the differences are not a
manifestation of evolution in the orbital properties at infall.
Rather, they must reflect processes that occur inside the host
halo. Using the average mass assembly histories of dark matter haloes
(see van den Bosch \etal 2014), we infer that the average halo with
mass in the range $10^{12} \Msunh$ to $10^{15} \Msunh$ grows its
virial radius by about $20\pm 5$ percent since $z = 0.25$. This effect
by itself only explains about half of the difference in the
$r_\rmc(E)/r_{\rm vir}$ distributions of A and E subhaloes, and has a
negligible impact on the $\eta$-distribution (at least if the halo
growth is adiabatic). Dynamical friction can drastically lower the
orbital energy, but only for the most massive subhaloes, i.e., those
with $m_{\rm acc}/\Mhalo \gta 0.1$ (see discussion in van den Bosch
\etal 2016). As is evident from Fig.~\ref{fig:massfunc}, the mass
function of subhaloes at accretion has a much larger fraction of
massive subhaloes than that of the surviving or ejected
subhaloes. Hence, dynamical friction plays an important role in
reducing the orbital energies, and thus in lowering the average
$r_\rmc(E)/r_{\rm vir}$. It also explains, at least partially, why
subhaloes that are about to be ejected have larger circularities than
orbits at accretion. As shown in van den Bosch \etal (1999), during
peri-centric passage dynamical friction causes $\eta$ to increase.
Note that the inverse is true at apo-center, where dynamical friction
causes $\eta$ to decrease again resulting in little net change in
$\eta$ when averaged over an entire radial orbit. However, since
subhaloes between accretion and ejection only experience a {\it
  peri}-centric passage, dynamical friction will cause a net shift in
the distribution of orbital circularities to larger values (more
circular orbits). The third effect that causes a large change in
orbital properties between accretion and ejection is subhalo
disruption. As we will show in \S\ref{sec:DPI} below, a significant
fraction of newly accreted subhaloes is disrupted during their first
peri-centric passage or shortly thereafter. Since subhaloes that are
disrupted are preferentially on radial orbits, this helps to explain
the pronounced deficit of small-$\eta$ orbits among the population of
E (and T) subhaloes.  To conclude, halo growth, dynamical friction and
tidal disruption all play an important role in causing a dramatic
change in the distribution of orbital properties of subhaloes between
accretion and ejection.

As mentioned above, 59 percent of the surviving, present-day subhalo
population in our sample have an apo-center $r_\rma > r_{\rm vir}$,
and are therefore expected to experience ejection.  At first, this may
seem inconsistent with the fact that the ratio $f_\rmE/f_\rmA$,
averaged over host haloes with $10^{12.5} \leq
\Mhalo/(h^{-1}\Msun)\leq 10^{14.5}$, is $1.0 \pm 0.02$, where the
error reflects the scatter among different output-pairs separated by
$\Delta t = 170\Myr$. However, the interpretation of the ratio
$f_\rmE/f_\rmA$ is complicated by the fact that there are, on average,
$\sim 3$ Gyrs between accretion and ejection during which the accretion
rate of a host halo may undergo appreciable changes. Hence, one ought
to compare the present-day ejection fraction to the accretion fraction
at the (average) redshift at which those subhaloes were accreted ($z
\simeq 0.25$).  In general, the accretion rate of dark matter haloes
declines with time (e.g., Neistein \& Dekel 2008; Fakhouri, Ma \&
Boylan-Kolchin 2010; van den Bosch \etal 2014), which implies that
the fraction of accreted subhaloes that is ejected is actually smaller
than unity. Assuming that $f_\rmA \propto \dot{M}_\rmh/M_\rmh$, and
using that $\dot{M}_\rmh/M_\rmh \propto (1+z)^{2.25}$ (Dekel \etal
2009), we have that $f_\rmA(z=0.25) \simeq 1.65 f_\rmA(z=0)$. Hence,
$f_\rmE(z=0)/f_\rmA(z=0.25) \simeq 0.6$, which is in excellent
agreement with the finding that 59 percent of present-day subhaloes
have $r_\rma > r_{\rm vir}$. The implication is that roughly 40
percent of the subhaloes that were accreted around $z=0.25$ do not
experience ejection; either because they were disrupted during their
first peri-centric passage, or because of changes in their orbital
properties.

The lower panel in the middle column of Fig.~\ref{fig:AER} shows that
an average subhalo at ejection has already experienced significant
mass loss since accretion; the average $m/m_{\rm acc}$ for E subhaloes
in our sample is $0.77$. This is not enough, however, to explain the
large differences in the subhalo mass functions of A and E subhaloes
evident from the upper panels of Fig.~\ref{fig:massfunc}. Instead, as
is evident from the lower panels, the E subhaloes already had a very
different mass function at accretion. This clearly indicates that the
subhaloes that do not experience ejection are predominantly the most
massive subhaloes, which are the ones that experience the largest
amount of dynamical friction.

Comparing the ejection and re-accretion fractions, we find that
$f_\rmR/f_\rmE = 0.83\pm 0.03$. If we assume that the ejection
fraction evolves in the same way as the accretion fraction, i.e.,
$f_\rmE \propto \dot{M}_\rmh/M_\rmh$, then the relevant ratio is
$f_\rmR(z=0)/f_\rmE(z=z_\rmE) \simeq 0.83 (1+z_\rmE)^{-2.25}$ where
$z_\rmE$ is the average ejection redshift for subhaloes that are
re-accreted at $z=0$.  This indicates that a significant fraction of
subhaloes that is being ejected will not be re-accreted. We emphasize
that this is not due to subhaloes having velocities that exceed the
escape speed\footnote{Some subhaloes that are ejected from their host
  have experienced a three-body interaction that causes the subhalo to
  become unbound, and thus to escape without being re-accreted (e.g.,
  Sales \etal 2007).}. After all, subhaloes on unbound orbits have
been removed from our analysis. Hence, the fact that
$f_\rmR(z=0)/f_\rmE(z=z_\rmE) < 1$ indicates that external tidal
forces from neighboring haloes can have an appreciable impact on their
orbit. In the extreme case, this may result in the subhalo being
captured by another host halo, thus giving rise to subhalo `exchange'
(as reflected by the X$^{+}$ and X$^{-}$ channels). As is evident from
Fig.~\ref{fig:AER}, R subhaloes have orbital properties that are
significantly different from those of E subhaloes, supporting the
notion that external tides influence the orbital energy and/or angular
momentum of subhaloes during their excursion outside of their
host. The net effect is that R subhaloes are on more radial orbits
than their E counterparts.

\subsection{Stripping and Merging}
\label{sec:SM}

Stripping, S, and merging, M, are each others inverse. Subhaloes in S
transit from being a higher-order subhalo at $t_\rmi$ to a first-order
subhalo at $t_\rmf$, while those in M transit from a first-order
subhalo at $t_\rmi$ to a higher-order subhalo at $t_\rmf$. In
principle, there are three effects that can contribute to stripping
and merging: First of all, S and M could represent true `physical'
processes. For stripping, this corresponds to a higher-order subhalo
being stripped off from its first-order parent halo by the tidal force
of either the host halo or another subhalo, such that it is no longer
bound to that parent.  For merging this corresponds to two first-order
subhaloes, initially unbound to each other, merging together such that
the less massive one ends up on a bound orbit inside the more massive
one. Secondly, merging followed by stripping can also be a
manifestation of the ejection and re-accretion of higher-order
subhaloes; similar to first-order subhaloes, 59 percent of which are
on orbits with apo-centers that fall outside of their host halo,
second-order subhaloes can be on orbits that take them temporarily
outside of their (first-order) parent subhalo. This manifests itself
as stripping followed by merging. And finally, stripping and merging
may be manifestations of penetrating encounters, whereby two subhaloes
that are unbound to each other, and with radial extents $R_1$ and
$R_2$, have a close encounter with an impact parameter $b < {\rm
  MAX}[R_1,R_2]$.

In order to discriminate between these three possibilities, we now
focus on the demographics of the S and M channels.  Intriguingly, the
fractions of subhaloes evolving, at any given instant, along the S and
M channels are virtually identical. For the subhaloes in our sample we
find that $f_\rmS/f_\rmM = 0.99 \pm 0.02$, where the average and
standard deviation are taken over all 171 output pairs. Not only are
their fractions identical, S and M subhaloes also have very similar
mass functions (see Fig.~\ref{fig:massfunc}), are on virtually
identical orbits, and have indistinguishable properties (see
Fig.~\ref{fig:SM}).  Comparing the properties of S and M subhaloes to
those of all subhaloes, it is clear that S and M subhaloes are fairly
representative of the full subhalo sample, except for their
distribution of halo-centric radii, $r/r_{\rm vir}$, which is clearly
skewed towards smaller values. This indicates that stripping and
merging preferentially occur closer towards the center of the host
halo. S and M subhaloes also seem to avoid the most circular orbits
(i.e., those with $\eta$ close to unity).
\begin{figure}
\includegraphics[width=\hssize]{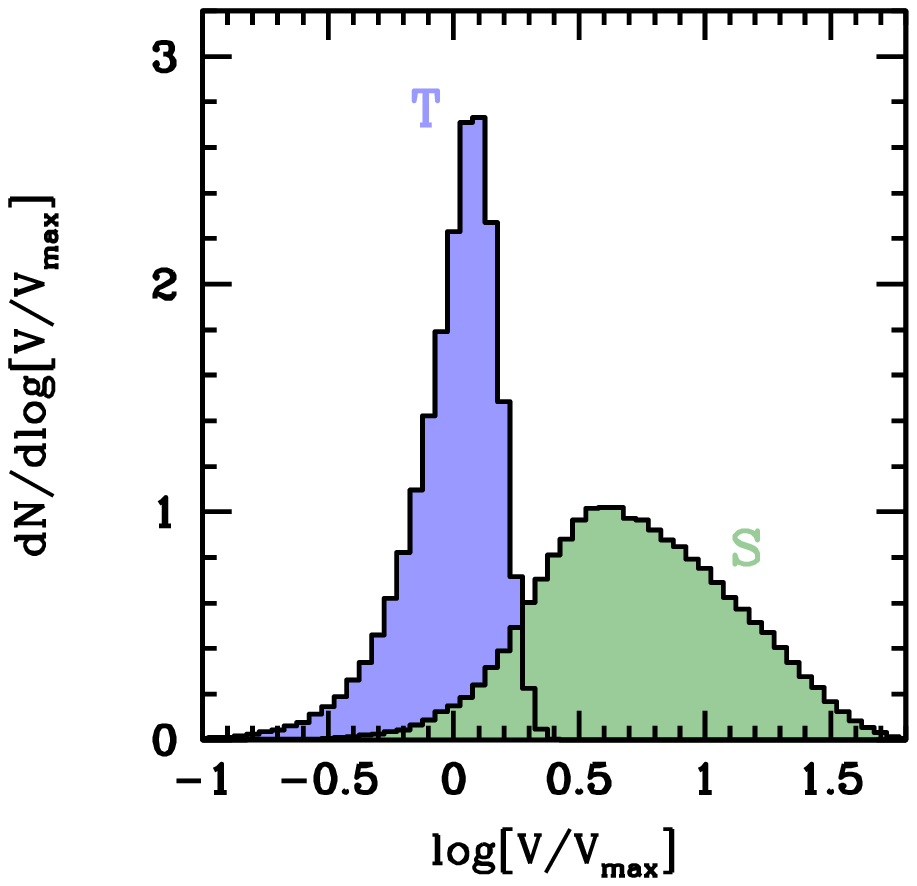}
\centering
\caption{The green histogram shows the distribution of $\log[V/V_{\rm
      max}]$, where $V$ is the velocity of a second-order S subhalo
  (about to be stripped) with respect to its first-order parent halo,
  and $V_{\rm max}$ is the maximum circular velocity of that parent
  halo. For comparison, the blue histogram shows the distribution of
  $\log[V/V_{\rm max}]$ for T subhaloes with $V$ the velocity of the
  subhalo with respect to its host halo, and $V_{\rm max}$ the maximum
  circular velocity of that host halo. Since, by construction, all T
  subhaloes are bound, this is representative of the $V/V_{\rm max}$
  distribution for a bound population. Clearly then, the vast majority
  of S subhaloes are not bound to their parent halo, indicating that S
  subhaloes are undergoing a penetrating encounter with another
  subhalo.}
\label{fig:SMvel}
\end{figure}

The fact that subhaloes along the S and M channels are virtually
indistinguishable, and that stripping and merging occur at virtually
identical rates, makes it extremely unlikely that the S and M channels
represent physical stripping and merging of higher-order subhaloes.
This is also supported by Fig.~\ref{fig:time}, which shows that
$f_\rmS$ and $f_\rmM$ do not increase linearly with $\Delta t$, as
would be expected if stripping and merging are independent physical
processes. Rather, we find S and M to be strongly correlated
temporally.  A subhalo that merges between time steps $t_{n-1}$ and
$t_n$, has a probability of being stripped between time steps $t_n$
and $t_{n+1}$ that is 6 times higher than for an average subhalo. As a
consequence, we find that only 50 percent of subhaloes that merge
between two outputs remain merged (i.e., are not stripped) for more
than $\sim 0.3\Gyr$. This implies that the S and M channels do not
correspond to actual stripping and merging (as defined above) of
higher-order subhaloes. In addition, it also makes it unlikely that
the majority of stripping and merging events are manifestations of the
orbits of second-order subhaloes. After all, $0.3\Gyr$ is about an
order of magnitude shorter than the typical orbital time between
accretion (which manifests itself as merging) and ejection (which
manifests itself as stripping).

Hence, the S and M channels most likely correspond to penetrating
encounters between subhaloes. This is corroborated by
Fig.~\ref{fig:SMvel}, which compares the distributions of $V/V_{\rm
  max}$ for (second-order) S subhaloes (green histogram), and
(first-order) T subhaloes (blue histogram). Here $V$ is the velocity
of the subhalo with respect to its parent/host halo, whose maximum
circular velocity is given by $V_{\rm max}$. As mentioned in
\S\ref{sec:SIM}, we exclude from our analysis any first-order subhalo
that is not bound to its host halo.  Consequently, the blue histogram,
which cuts-off sharply around $V/V_{\rm max} \sim 2$, is
representative of a population of bound subhaloes. The $V/V_{\rm max}$
distribution for S subhaloes is very different. It has a median of
$V/V_{\rm max} = 5.2$, with 87.4 percent having $V/V_{\rm max}>2$
(compared to 0.6 percent for T subhaloes); clearly, the vast majority
of (second-order) S subhaloes are not bound to their (first-order)
parent subhalo. We therefore conclude that the S and M channels are
manifestations of high-speed penetrating encounters. In
\S\ref{sec:enc} we show that the inferred rate of penetrating
encounters is remarkably high, which may have important ramifications
for the evolution of subhaloes.
\begin{figure*}
\includegraphics[width=\hdsize]{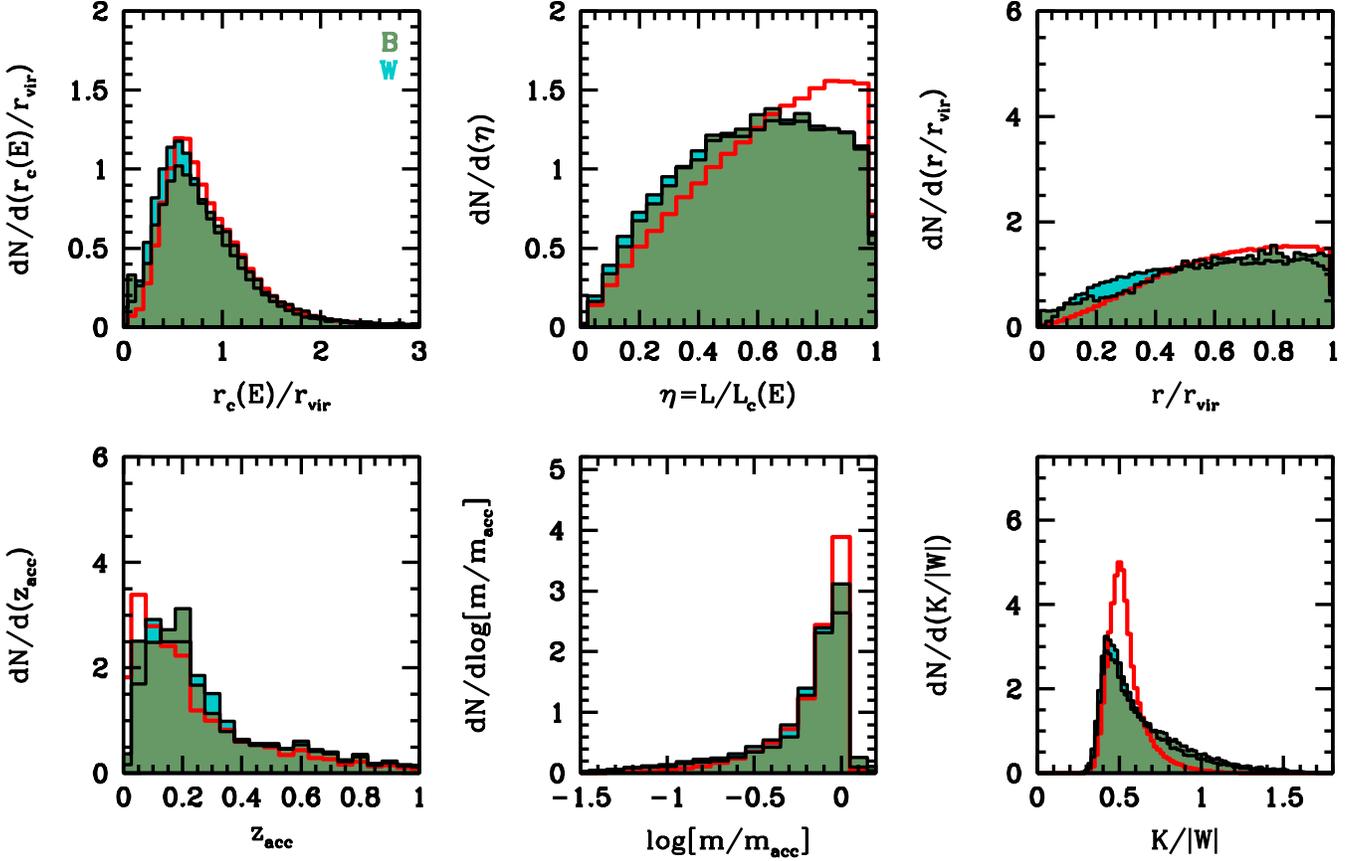}
\centering
\caption{Same as Fig.~\ref{fig:AER}, but for channels W and B.}
\label{fig:WB}
\end{figure*}

\subsection{Withering and Blossoming}
\label{sec:WB}

Withering subhaloes, W, are defined as subhaloes whose mass drops
below the 50 particle limit during the interval $\Delta t$. Blossoming
is the inverse of withering, and B subhaloes therefore have their mass
increase from below $\mmin$ to above $\mmin$. In general, one expects
subhaloes to loose mass, and not gain mass, and thus for the B channel
to be empty. Yet, we find that, in our fiducial time interval of
$\Delta t \simeq 170\Myr$, about 0.6 percent of all subhaloes in our
sample are blossoming. For comparison, in the same time interval
roughly 2.3 percent of subhaloes wither.

Similar to S and M subhaloes, W and B subhaloes are strongly
correlated temporally.  Subhaloes that just blossomed are about
$50\times$ more likely to wither within the next $42\Myr$ than an
average subhalo, while subhaloes that withered are $\sim 60\times$
more likely to have blossomed during the previous time step of $42
\Myr$. This explains why the normalized, fractional rates for the W
and B channels, shown in Fig.~\ref{fig:time} decrease with increasing
$\Delta t$ (i.e., the fractions $f_\rmW$ and $f_\rmB$ do not increase
linearly with $\Delta t$). Half of all subhaloes that blossom will
wither again in $\sim 100\Myr$, while roughly 85 percent of withering
subhaloes are gone from the sample indefinitely. The remaining 15
percent blossom again within about $100\Myr$.  All of this suggests
that blossoming is mainly a manifestation of the mass histories of
subhaloes being erratic, such that subhaloes with a mass close to
$\mmin$ repeatedly `scatter' above and below this limit (see
\S\ref{sec:massloss} below).  If we interpret all blossoming as merely
temporal fluctuations in $m(t)$, then we can correct the withering
fraction by simply subtracting the blossoming fraction. This implies
an effective, fractional withering rate of 0.10 Gyr$^{-1}$ (i.e., per
Gyr roughly 10 percent of all subhaloes wither, without ever blossoming
again).

As is evident from Fig.~\ref{fig:WB}, withering and blossoming
subhaloes have properties that are virtually indistinguishable from
each other, and are very representative of an average subhalo.  In
particular, blossoming and withering do not preferentially occur along
a particular subset of orbits, although they appear to `avoid' the
most circular orbits. There is also some indication that W subhaloes
are, on average, at smaller halo-centric distances, consistent with
the notion that mass loss is more prevalent closer to peri-center.  In
addition, W and B subhaloes have virial ratios, $K/|W|$, that are
slightly more skewed towards larger values. 

Finally, the right-hand panels of Fig.~\ref{fig:massfunc} show the
cumulative mass functions of W and B subhaloes.  Whereas withering
subhaloes always have masses close to $\mmin$, blossoming subhaloes
can have masses that are quite substantial.  In fact roughly 2 (0.2)
percent of all B subhaloes in our sample have more than 100 (250)
particles. If indeed, as argued above, blossoming subhaloes are a
manifestation of noise in the subhalo mass assignment, this indicates
that the error in the assigned subhalo mass can be disturbingly
large. As we demonstrate in \S\ref{sec:massloss} below, this
is indeed the case.
\begin{figure*}
\includegraphics[width=\hdsize]{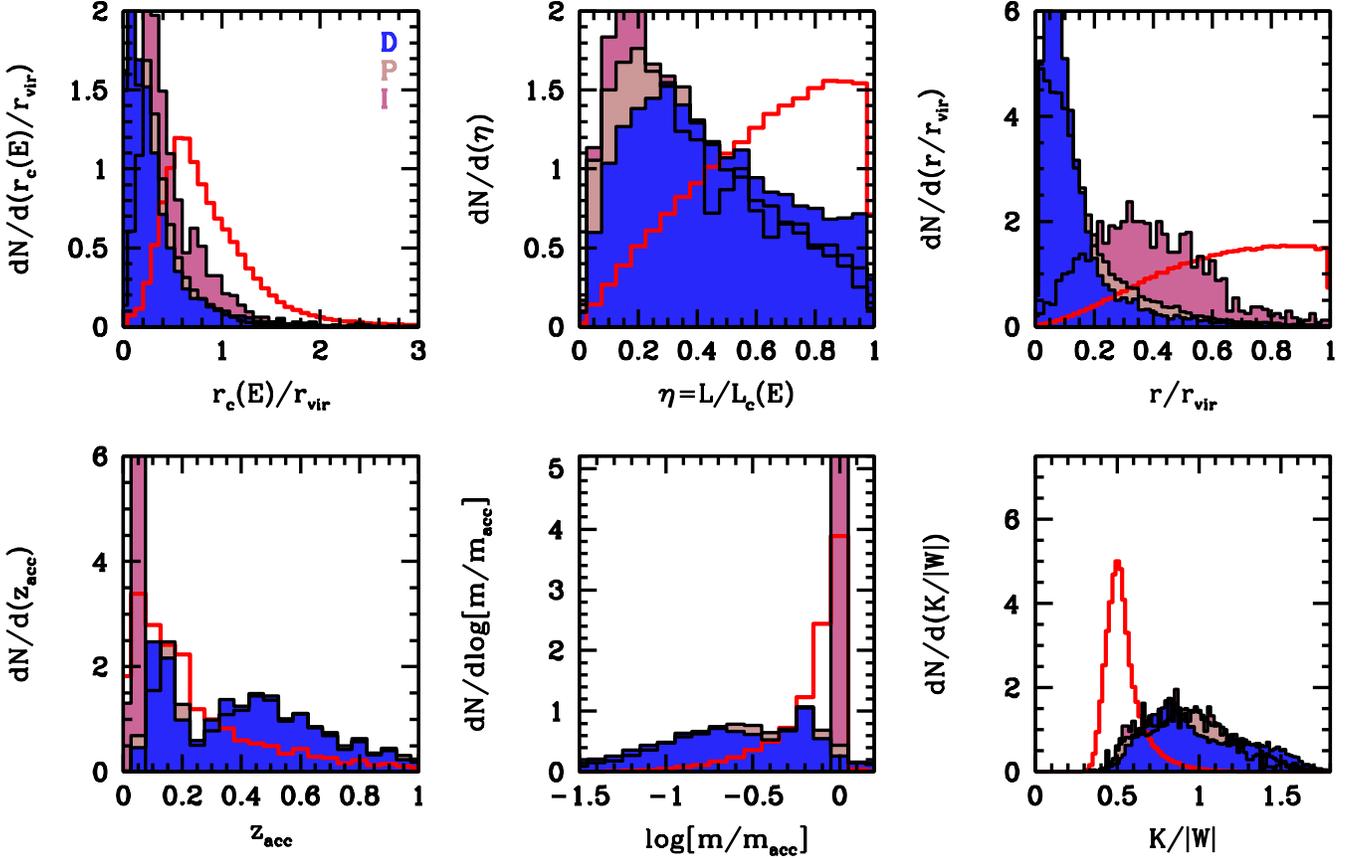}
\centering
\caption{Same as Fig.~\ref{fig:AER}, but for the disruption (\cD) and
  immaculate (\cI) channels, as well as for phantom subhaloes (\cP).}
\label{fig:DPI}
\end{figure*}

\subsection{Disruption, Phantoms and Immaculates}
\label{sec:DPI}

We now turn our attention to subhalo disruption. D subhaloes are
defined as subhaloes that at $t_\rmi$ have a mass $m > \mmin$, but
that are not the most massive progenitor of their descendant halo.  As
discussed in \S\ref{sec:trees}, in the output before their
disappearance from the halo catalogs, they experience a tidal
acceleration $|\calT| > 0.4 \kms \Myr^{-1}$ comoving Mpc$^{-1}$;
subhaloes that disappear from the catalogs but that do not meet this
tidal criterion are considered `merger fluctuations' and are removed
from the catalogs, together with all their progenitors at all previous
time steps.

As is evident from Fig.~\ref{fig:fraction_mass} disruption rates are
higher in less massive host haloes. This is consistent with
expectations; subhaloes in lower mass hosts were accreted earlier,
have experienced more mass loss on average, and are therefore more
likely to be susceptible to disruption.  In addition, less mass host
haloes are more centrally concentrated, and therefore cause stronger
tidal shocking. This latter effect is somewhat counter-balanced,
though, by the fact that less massive hosts will accrete less massive
subhaloes, which themselves are more concentrated, and hence more
resilient to disruption.

Fig.~\ref{fig:DPI} compares the properties of D subhaloes to those of
immaculate subhaloes (I) and phantom subhaloes (P).  Note that P
subhaloes are not part of our complete set of evolution channels.
Rather, phantoms are almost exclusively part of the T channel.  All in
all these D and P subhaloes have very similar properties. Their orbits
are far more bound and radial than for an average subhalo (i.e., they
have smaller $r_\rmc(E)/r_{\rm vir}$ and smaller $\eta$). Both D and P
subhaloes are located at small halo-centric radii (54\% of D subhaloes
have $r/r_{\rm vir} < 0.1$, compared to 1\% for T subhaloes), have
elevated virial ratios (the median $K/|W| = 0.924$ for D subhaloes,
compared to 0.521 for T subhaloes), and have clearly experienced more
mass loss than an average subhalo.  All of this is consistent with
disruption being caused by the strong tidal field of the host halo
near the orbit's peri-center.

The distribution of accretion redshifts for D subhaloes shows a
pronounced absence at $z_{\rm acc}=0$ and around $z_{\rm acc} \simeq
0.25$.  Instead, D subhaloes have preferentially been accreted around
$z_{\rm acc} \sim 0.15$ and $0.45$. As shown in van den Bosch \etal
(2016), subhaloes accreted at these redshifts are experiencing, on
average, their first and second peri-centric passages around $z=0$,
whereas subhaloes accreted at $z \sim 0.25$ are currently experiencing
their first apo-centric passage. Hence, this distribution of accretion
redshifts is exactly what one expects if disruption occurs
preferentially near peri-center.

The fact that phantom subhaloes have properties that are almost
indistinguishable from D subhaloes raises the concern that the latter
are actually phantoms, i.e., subhaloes that are not disrupted, but
that temporarily disappear from the halo catalogs because of issues
with the subhalo finder. Recall that a subhalo can only be assigned
`phantom status' for four consecutive time steps. Hence, if a subhalo
fails detection in five or more consecutive time steps, it will show
up in our sample as a disrupted subhalo. However, if disrupted
subhaloes are really phantoms that are not traced long enough, then
the majority of D subhaloes should have a corresponding I subhalo.
However, the fraction of immaculates is much lower than that of D
subhaloes ($f_\rmI/f_\rmD \lta 0.1$), indicating that at most a small
subset of D subhaloes can be mis-classified phantoms (they don't
really disrupt, they merely temporarily disappear from the halo
catalogs). This is also supported by the left-hand panels of
Fig.~\ref{fig:massfunc}, which show that the mass function of D
subhaloes is very different from that of I subhaloes.

However, there is one example where phantoms are clearly contributing
significantly to the disruption channel, which is at $z \lta
0.012$. As shown in Fig.~\ref{fig:fraction_redshift}, disruption seems
to become somewhat more prevalent at these low redshifts. However,
this is entirely an artifact of contamination by phantom haloes, and
arises from the fact that the simulation is only run to $z=0$. Any
subhalo that becomes a phantom in one of the final three outputs prior
to $z=0$ and is set to `resurface' in the future (not covered by the
simulation), ends up being (erroneously) identified as a disrupted
subhalo. This explains the slight increase in the fraction of D
subhaloes for $z \lta 0.012$.

As argued above, the fact that the fraction of immaculates is so much
smaller than that of disrupted subhaloes indicates that the majority
of disrupted subhaloes cannot be misclassified phantoms. However, it
is still possible that the vast majority of immaculates are actually
phantoms that simple weren't traced long enough. This is, in fact,
supported by the fact that immaculates have orbital properties that
are extremely similar to those of phantoms. As is evident from the
upper right-hand panel of Fig.~\ref{fig:DPI}, immaculates are located
at somewhat larger halo-centric radii than D and P subhaloes (but
still much closer to the center than an average subhalo). This is
exactly what is expected if immaculates are actually phantoms that
resurface more than four time steps after they disappear, when
sufficient time has passed since the last peri-centric passage.
Therefore, we suspect that the majority of immaculates are actually
phantoms that have been traced for an insufficient amount of time, and
that their fraction will become even more insignificant if the merger
tree algorithm would allow the creation of phantoms for more than
four successive time steps.
\begin{figure*}
\includegraphics[width=0.9\hdsize]{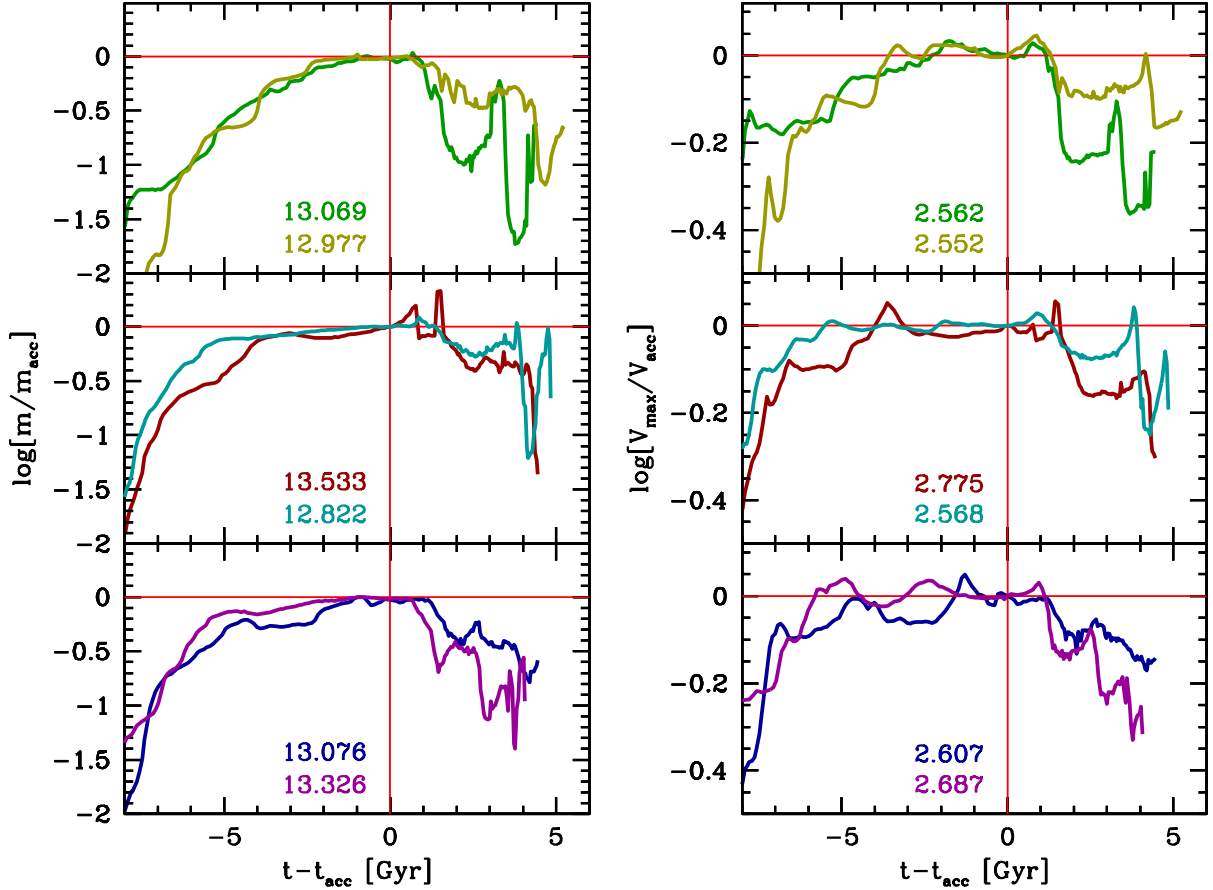}
\centering
\caption{Mass (left) and $V_{\rm max}$ (right) histories for 6
  subhaloes randomly selected from the Bolshoi simulation. To avoid
  crowding, we only show two histories per panel. The vertical, red
  line marks the epoch of accretion, $t_{\rm acc}$, after which the
  halo becomes a subhalo. The corresponding values of $\log[m_{\rm
      acc}/(h^{-1}\Msun)]$ and $\log[V_{\rm acc}/(\kms)]$ are
  indicated in each panel. Note the large fluctuations in mass and
  maximum circular velocity that the subhaloes experience after
  accretion, which are very different from what is seen in analytical
  models and idealized simulations.}
\label{fig:masshis}
\end{figure*}

\section{The Tidal Evolution of Dark Matter Subhaloes}
\label{sec:evolution}

While orbiting their host halo, subhaloes experience dynamical
friction, tidal stripping, and tidal heating during peri-centric
passages and impulsive encounters with other subhaloes. All of these
processes contribute to subhaloes loosing mass and/or being disrupted.
This section takes a closer look at the demographics of tidal
stripping and tidal disruption in the Bolshoi simulation. We start in
\S\ref{sec:massloss} with a detailed examination of the tidal
evolution of T subhaloes. We show that their mass and maximum circular
velocity histories are extremely erratic, something that is difficult
to reconcile with physical expectations. In \S\ref{sec:enc} we derive
an expression for the average time between penetrating encounters
among subhaloes. Finally, in \S\ref{sec:disruption} we show that there
are three distinct populations of disrupting subhaloes; a population
that `withers' below the mass resolution limit of the simulation, a
population that `merges' with the host, largely driven by dynamical
friction, and a population that seems to abruptly disintegrate,
something that again is difficult to reconcile with physical
expectations.

\subsection{Tidal Stripping}
\label{sec:massloss}

As shown in Jiang \& van den Bosch (2016a), the (orbit-averaged) mass
loss rate of dark matter subhaloes (in Bolshoi) is well described by
\begin{equation}\label{massloss}
{\rmd m \over \rmd t} = -0.86 {m \over \tau_{\rm dyn}(t)} \left( {m
  \over M}\right)^{0.07}
\end{equation}
where $\tau_{\rm dyn}(t)$ is the dynamical time of the host halo at
time $t$, and $m$ and $M$ are the instantaneous masses of the subhalo
and host halo, respectively (see also van den Bosch \etal 2005;
Giocoli \etal 2010). The average half-mass time at $z=0$ for a subhalo
ranges from $\sim 2.5 \,\Gyr$ for the most massive subhaloes to $>
5.0 \,\Gyr$ for subhaloes that at accretion have a mass $m < 10^{-5}
M$. On average, subhaloes accreted at $z_{\rm acc} \sim 0.32$ ($\sim
1.0$) have lost 50 (90) percent of their accretion mass at the
present. For comparison, the maximum time interval examined in this
study is only $0.81\Gyr$, and subhaloes, on average, only loose a few
percent of their mass over this period.

Note that Eq.~(\ref{massloss}) describes the orbit-averaged mass loss
rates, averaged over the entire distribution of orbits and orbital
phases.  In practice, because of the broad distributions of orbital
properties and halo concentrations, one expects a fair amount of
scatter in the instantaneous mass loss rates. In addition, analytical
calculations of tidal stripping indicate that subhaloes loose the
majority of their mass during the short periods associated with
peri-centric passages, giving rise to `stair-case' like behavior of
$m(t)$ (e.g., Taylor \& Babul 2001, 2004; Taffoni \etal 2003). This is
indeed the behavior seen in idealized simulations of a single $N$-body
subhalo orbiting within a fixed, analytical host halo potential (e.g.,
Hayashi \etal 2003; Pe\~narrubia \etal 2010). However, as we
demonstrate next, this is not at all what is seen in the Bolshoi
simulation.
\begin{figure*}
\includegraphics[width=0.95\hdsize]{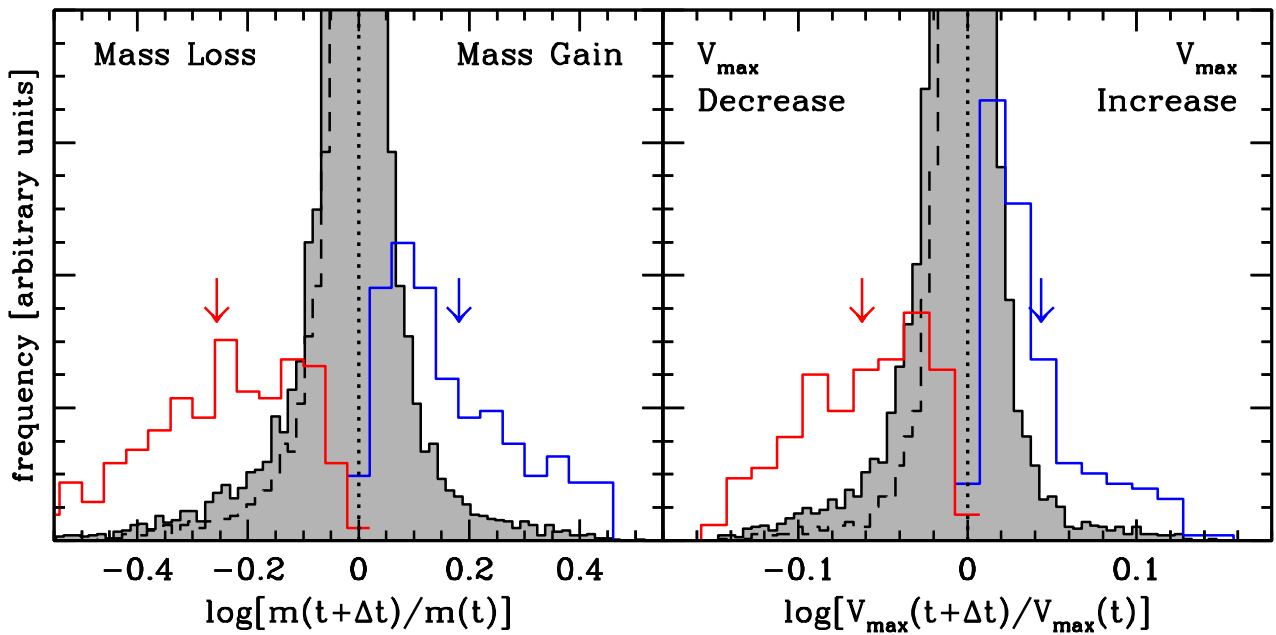}
\centering
\caption{{\it Left-hand panel:} The shaded histogram plots the
  distribution of $\log[m(t+\Delta t)/m(t)]$ for all T-subhaloes with
  present day mass $m_0 \geq 10^{12} h^{-1}\Msun$ and for all
  time-steps $t \rightarrow t + \Delta t$ with $\Delta t = 42 \Myr$ in
  which the halo is a subhalo. Positive and negative values correspond
  to mass gain and mass loss, respectively, as indicated.  The dashed
  line at $\log[m(t+\Delta t)/m(t)] < 0$ is the mirror-reflection of
  the histogram for $\log[m(t+\Delta t)/m(t)] > 0$, and is shown to
  highlight the high level of symmetry in the distribution. The red
  and blue histograms indicate the distributions of the minimum and
  maximum $\log[m(t+\Delta t)/m(t)]$, respectively, for individual
  subhaloes. The vertical arrows indicate the corresponding medians.
  {\it Right-hand panel:} same as left-hand panel, but for
  $\log[V_{\rm max}(t+\Delta t)/V_{\rm max}(t)]$. See text for
  detailed discussion.}
\label{fig:extreme}
\end{figure*}

Fig.~\ref{fig:masshis} shows examples of the mass (left-hand panels)
and $\Vmax$ (right-hand panels) histories for 6 randomly selected
subhaloes in the Bolshoi simulation. These were selected to have a
mass at $z=0$ in excess of $10^{12} \Msunh$ (i.e., more than 7400
particles).  The histories are normalized to the mass and maximum
circular velocity at the epoch of accretion, the logarithmic values of
which are indicated in each panel. Note the `noisy', jagged appearance
of both $m(t)$ and $\Vmax(t)$, which is very different from the
`stair-case like' behavior in analytical models and idealized
simulations. It is well known that assigning masses to subhaloes can
be tricky, especially if the subhalo is deeply embedded in the host
halo (e.g., Muldrew, Pearce \& Power 2011; Knebe \etal 2011; Han \etal
2012). In particular, it is extremely difficult to identify the outer
boundary of the halo, which results in uncertainties in the assigned
mass. However, the fact that the evolution in $\Vmax$, which is a measure
for the depth of the central potential well, is equally jagged as
that in mass, suggests that this is not merely an issue related to
identifying the subhalo's outer boundary.

The grey histograms of Fig.~\ref{fig:extreme} plot the distributions
of $\log[m(t+\Delta t)/m(t)]$ (left-hand panel) and $\log[V_{\rm
    max}(t+\Delta t)/V_{\rm max}(t)]$ (right-hand panel) for all
subhaloes with present day mass $m_0 \geq 10^{12} h^{-1}\Msun$ and for
all time-steps $t \rightarrow t + \Delta t$ with $\Delta t = 42 \Myr$
in which the halo is a subhalo. Positive and negative values
correspond to an increase and decrease, respectively, in $m$ or
$\Vmax$.  Note that the distributions are remarkably symmetric
indicating that roughly half the time subhaloes actually {\it gain}
mass, or increase their maximum circular velocity, at least according
to the \Rockstar halo finder. More precisely, 40\% (46\%) of the time
steps result in an {\it increase} of $m$ ($\Vmax$). The dashed lines
at the negative sides of the distributions are the mirror-reflections
of the histograms at the positive sides, and are shown to highlight
this high level of symmetry in the distributions. In the case of
subhalo mass, there is only a slight excess of mass loss over mass
gain, and it is this slight excess that is responsible for the overall
net mass loss experienced by subhaloes. Once again, this is very
different from the stair-case-like behavior discussed above, for which
$\log[m(t+\Delta t)/m(t)]$ only takes on negative values. For each
subhalo we have determined both the minimum and maximum
$\log[m(t+\Delta t)/m(t)]$ along their entire mass histories, the
distributions of which are indicated by the red and blue histograms in
Fig.~\ref{fig:extreme}, respectively. The vertical arrows indicate the
corresponding medians. The median, extremal mass gain is
$\log[m(t+\Delta t)/m(t)] = 0.18$, indicating that an average subhalo
in the Bolshoi simulation, at some point after its accretion, in a
time step of only $42 \Myr$, increases its mass by 50 percent. Note
also that the extrema-distribution extends well beyond
$\log[m(t+\Delta t)/m(t)] = 0.3$; roughly 19 percent of all subhaloes
experience time steps in which they more than double their own mass.
The results for $V_{\rm max}$, shown in the right-hand panel, are
qualitatively similar, with individual subhaloes revealing a median,
extremal increase in $V_{\rm max}$ of a factor 1.1.

Are these jagged $m(t)$ and $\Vmax(t)$ histories real, or are these
artifacts of the simulations and/or the subhalo finder. One way in
which a subhalo can fluctuate in mass is by experiencing a penetrating
encounter with another subhalo (halo masses in \Rockstar are inclusive
of all substructure). As discussed in \S\ref{sec:SM}, such penetrating
encounters manifest themselves via the stripping (S) and merging (M)
channels, and happen frequently. Since we only focus on first-order
subhaloes, a penetrating encounter can at most boost the mass by a
factor of two. Such equal-mass encounters, however, are extremely
rare. By examining the mass ratios of subhaloes belonging to the
M-channel, we find that only 0.97 (13.6) percent of the subhaloes that
`merge' contribute more than 10 (1) percent of mass to the subhalo
into which they merge. Hence, penetrating encounters only make a very
small contribution to the jaggedness of $m(t)$. Another way in which
subhaloes can grow in mass, in principle, is by accreting smooth
matter from their surroundings.  However, as discussed and
demonstrated in Han \etal (2012), subhaloes do not accrete any
significant amount of matter from their host halo.

\begin{figure*}
\includegraphics[width=\hdsize]{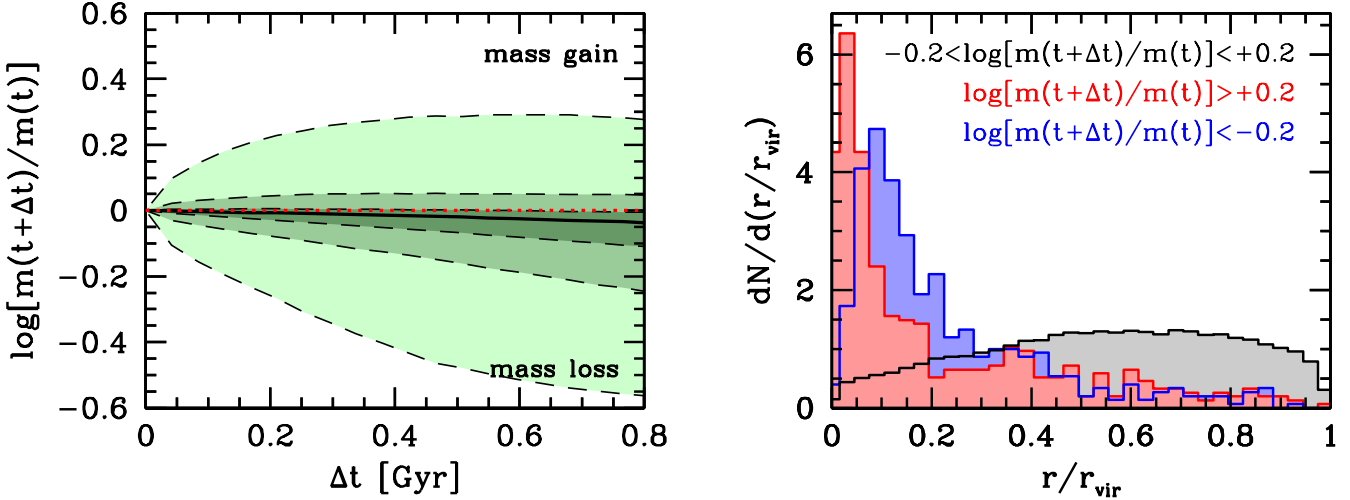}
\centering
\caption{{\it Left-hand panel:} The ratio $m(t+\Delta t)/m(t)$ as
  function of $\Delta t$ for all T subhaloes in our sample with $m_0 >
  10^{12} h^{-1} \Msun$. The thick, solid black line indicates the
  median, while the shaded regions mark the 68, 95 and 99
  percentiles. The red, dotted line indicates $m(t+\Delta t)/m(t) = 1$
  and is shown for comparison.  As in Fig.~\ref{fig:extreme} positive
  and negative values of $\log[m(t+\Delta t)/m(t)]$ correspond to mass
  gain and mass loss, respectively, as indicated. {\it Right-hand
    panel:} The distribution of halo-centric radii where subhaloes
  experience different amounts of mass loss or gain, as
  indicated. Note that episodes of extreme mass loss and mass gain
  preferentially occur at small halo-centric radii.}
\label{fig:massloss}
\end{figure*}

In order to gain some insight into the origin of the irregular, jagged
behavior of $m(t)$ and $\Vmax(t)$, the left-hand panel of
Fig.~\ref{fig:massloss} plots $\log[m(t+\Delta t)/m(t)]$ as function
of $\Delta t$. As in Fig.~\ref{fig:extreme}, we consider all T
subhaloes in our sample with a present-day mass $m_0 \geq 10^{12}
\msunh$. The thick, solid curve indicates the median, while the shaded
regions mark the 68, 95 and 99 percentile intervals. Over the
$0.8\Gyr$ covered by our study, the average T subhalo only looses
about 8 percent of its mass, but with a subhalo-to-subhalo variance
that covers the range $0.27 < m(t+\Delta t)/m(t) < 1.85$ (99\%
interval).  Most importantly, the positive (mass gain) wing of the
$m(t+\Delta t)/m(t)$ distribution increases with $\Delta t$, at least
up to $\Delta t \sim 0.4\Gyr$. This indicates that the episodes of
(strong) subhalo mass increase are coherent over time-scales of
$\Delta t \sim 0.4 \Gyr$; this coherence is also evident from the mass
histories shown in Fig.~\ref{fig:masshis}.

The right-hand panel of Fig.~\ref{fig:massloss} plots the
distributions of halo-centric radii within the host halo, $r/r_{\rm
  vir}$, at which the subhaloes experience time-steps of $\Delta t =
42 \Myr$ in which $\log[m(t+\Delta t)/m(t)] > +0.2$ (extreme mass
gain; red histogram), $\log[m(t+\Delta t)/m(t)] < -0.2$ (extreme mass
loss; blue histogram), and $-0.2 < \log[m(t+\Delta t)/m(t)] < +0.2$
(modest change in mass; black histogram).  Clearly, extreme mass loss
and gain occurs preferentially at small halo-centric radii. This is
expected for the mass loss, which, as discussed above, peaks during
peri-centric passage. However, the episodes of extreme mass gain,
which typically occur at even smaller halo-centric radii, are not
expected.  As shown in Han \etal (2012), subhaloes can re-accrete some
of their own matter that was stripped off at earlier times. In
particular, during a peri-centric passage a subhalo experiences a
drastic drop in mass, followed by a short period in which it
`re-accretes' some of this mass. Physically, this can come about
because the tidal stripping and shocking has pushed the subhalo out of
virial equilibrium; consequently, the subhalo undergoes
re-virialization, which can in principle re-bind some of the particles
that immediately after the impulsive shock have positive binding
energy. However, as discussed in van den Bosch \etal 2016, in prep.),
the magnitude of this physical `re-binding' is expected to be very
small (few percent increase in mass at most), much smaller than the
mass growth episodes evident in Fig.~\ref{fig:masshis}.

We are left to conclude that the jaggedness in $m(t)$ and $\Vmax(t)$
most likely reflects serious problems with the subhalo finder. As
demonstrated by Han \etal (2012), the detailed $m(t)$ in numerical
simulation during a peri-centric passage is extremely sensitive to
details regarding the unbinding algorithm used as part of the subhalo
finder: in particular, the AMIGA Halo Finder (AHF; Knollmann \& Knebe
2009) and \Rockstar typically yield larger `fluctuations' in $m(t)$
during peri-centric passage than the Hierarchical Structure Finder
(HSF; Maciejewski \etal 2009) or the Hierarchical Bound-Tracing (HBT;
Han \etal 2012) algorithm. Hence, we conclude that the fluctuations in
subhalo mass and maximum circular velocity reflect problems with the
\Rockstar subhalo finder, which become extreme whenever the subhalo is
located close to the center of its host halo. Similar studies are
required to examine how other subhalo finders fair in this respect.

Finally, although individual subhaloes may occasionally carry very
substantial errors in their instantaneous mass and/or maximum circular
velocity, in Appendix~\ref{App:convolution} we show that this does not
introduce an appreciable, systematic error in the inferred subhalo
mass or $\Vmax$ functions.

\subsection{Rate of Penetrating Encounters}
\label{sec:enc}

As we have seen in \S\ref{sec:SM}, the stripping and merging channels
are mainly manifestations of penetrating encounters among subhaloes.
Here a penetrating encounter is defined as having an impact parameter
$b < \max(R_1,R_2)$, with $R_1$ and $R_2$ the radii of the subhaloes
involved in the encounter. In this section we derive a simple estimate
for the rate of such penetrating encounters.

Let $\tau_{\rm enc}(m_0)$ be the mean time for a subhalo of mass $m_0$
between two penetrating encounters with subhaloes of mass $m >
m_0$. For a subhalo, the cross section for a penetrating encounter,
$\sigma$, is independent of its velocity with respect to the host
halo, $v$. Hence,
\begin{equation}
\tau_{\rm enc}(m_0) = {1 \over n \, \langle \sigma \rangle \, \langle v
  \rangle}\,,
\end{equation}
where $n$ is the number density of subhaloes with $m > m_0$.  As shown
by Gao \etal (2004) , the subhalo abundance {\it per unit halo mass}
is roughly $\rmd n/\rmd m \propto m^{-1.9}$, with a normalization that
is independent of host halo mass (see also van den Bosch \etal
2005). The interaction cross section $\sigma = \pi R^2$ where $R$ is
the radius of the more massive subhalo. Assuming that the size of a
subhalo of mass $m$ is proportional to $m^{1/3}$, we thus have that
$\sigma \propto m^{2/3}$. Averaging over all subhaloes of mass $m >
m_0$ one finds that
\begin{equation}
n \, \langle \sigma \rangle = {M \over V} \int_{m_0}^{M} {\rmd n \over \rmd m}
\sigma(m) \, \rmd m \propto m_0^{-0.23}\,,
\end{equation}
where we have used that, according to the halo virial relations, the
volume of a halo $V \propto M$, and we have assumed that $m_0 \ll M$.
Finally, using that $\langle v \rangle \propto M^{1/3}$, we obtain
that
\begin{equation}
\tau_{\rm enc}(m_0) \propto M^{-1/3} \, m_0^{0.23}\,.
\end{equation}
When considering all subhaloes of mass $m_0 > m_{\rm min}$, the average
time between penetrating encounters is simply
\begin{equation}\label{tauscaling}
\langle \tau_{\rm enc} \rangle \propto M^{-1/3} \, m_{\rm min}^{0.23}
\end{equation}

Let $f_\rmM$ be the fraction of subhaloes that evolve along the
merging channel in a time interval $\Delta t$. Since merging is merely
a manifestation of penetrating encounters with other, {\it more
  massive} subhaloes (see \S\ref{sec:SM}), we have that $\langle
\tau_{\rm enc} \rangle = \Delta t/f_\rmM$, as long as $\Delta t$ is
sufficiently small so that we can ignore multiple encounters per
subhalo (i.e., as long as $f_\rmM \ll 1$). Using the smallest
time-step available for the Bolshoi simulation ($\Delta t = 42 \Myr$),
we find that $f_\rmM$ increases from $\sim 0.006$ for $M = 10^{12}
\Msunh$ to $\sim 0.05$ for $M = 10^{15} \Msunh$. Note that this
scaling with host halo mass is in excellent agreement with our simple
estimate of Eq.~(\ref{tauscaling}), according to which $\langle
\tau_{\rm enc} \rangle \propto M^{-1/3}$. Using that $f_\rmM$
corresponds to a minimum subhalo mass of $m_{\rm min} = m_{50} = 6.75 \times
10^9 \Msunh$, we infer an average time in between two penetrating
encounters of
\begin{equation}\label{tau1enc}
\langle \tau_{\rm enc} \rangle \simeq (5 \pm 0.5) \, M_{12}^{-1/3} \,
\left({m_{\rm min} \over 10^9 \Msunh}\right)^{0.23}\,\Gyr\,,
\end{equation}
where $M_{12}$ is the host halo mass in units of $10^{12} \Msunh$.
The above can be recast as
\begin{equation}\label{tau2enc}
\langle \tau_{\rm enc} \rangle \simeq (3 \pm 0.3) \, M_{12}^{-0.1} \,
\left({m_{\rm min}/M \over 10^{-4}}\right)^{0.23}\,\Gyr\,.
\end{equation}
Note that both of these expressions are only valid at $z \sim 0$ 
and for $m_{\rm min}/M \ll 1$.

Comparing the mean free time, $\langle \tau_{\rm enc} \rangle$, to the
host halo's dynamical time $\tau_{\rm dyn} = \sqrt{3\pi/16 G
  \bar{\rho}} \simeq 3.1 \Gyr$ it is clear that penetrating encounters
of subhaloes with more massive subhaloes is extremely common (see also
Tormen \etal 1998). In particular, subhaloes with a mass $m = 10^8
\Msun$ orbiting inside a Milky-Way size host halo have on average
about one penetrating encounter with another, more massive subhalo per
dynamical time.  Since the majority of these encounters are high speed
(see Fig.~\ref{fig:SMvel}), it is clear that impulsive encounters
among subhaloes may play an important role in their evolution (see
also Moore \etal 1996).
\begin{figure}
\includegraphics[width=\hssize]{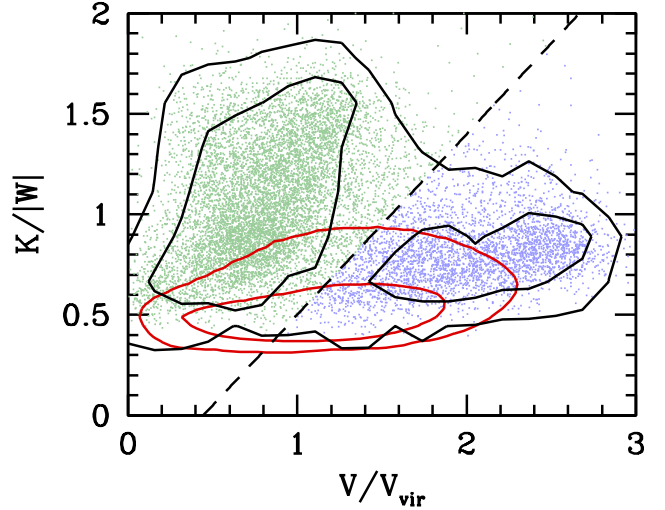}
\centering
\caption{The distribution of disrupting subhaloes in the parameter
  space of the virial ratio $K/|W|$ versus the ratio $V/V_{\rm vir}$,
  which is the instantaneous velocity of the subhalo with respect to
  its host halo in units of the host halo's virial velocity. The black
  contours indicate the 68 and 95 percentiles of this distribution,
  which is clearly bimodal. The dashed line marks the critical virial
  ratio $(K/|W|)_{\rm crit}$ (Eq.~[\ref{KWcrit}]), which is used to
  separate D-subhaloes into two categories: \Dm (green dots), which
  have a virial ratio larger than $(K/|W|)_{\rm crit}$, and \Ds (blue
  dots) for which $K/|W| < (K/|W|)_{\rm crit}$. For comparison, the
  red contours indicate the 68 and 95 percentiles of the distribution
  of T-subhaloes in $(K/|W|, V/V_{\rm vir})$-space.}
\label{fig:bimodal}
\end{figure}

\subsection{Tidal Disruption}
\label{sec:disruption}

The combined impact of tidal stripping and tidal heating can result in
the complete disruption of a subhalo. Indeed, one of the evolution
channels identified in this work is the disruption channel D. Although
disruption may appear insignificant, given that the fractional
contribution shown in \S\ref{sec:fractions} is only $\sim 0.5$
percent, be aware that this is the fractional contribution for a short
time interval of $\Delta t = 170\Myr$. As is evident from
Fig.~\ref{fig:time}, to good approximation $f_\rmD \propto \Delta t$,
which implies a disruption rate of $2.4$ percent per Gyr. It is
important to stress that this only accounts for subhaloes that at the
moment of disruption have a mass $m \geq m_{50}$. There are also
subhaloes that disrupt after first experiencing mass loss to (well)
below $m_{50}$. As discussed in \S\ref{sec:WB}, the effective rate at
which this occurs is given by the difference between the withering (W)
rate and the blossoming (B) rate, and is roughly 0.10 Gyr$^{-1}$.
Hence, if we include this effective withering as an additional
disruption channel, the total disruption rate in the Bolshoi
simulation amounts to $\sim 13$ percent per Gyr.

Indeed, as shown in Jiang \& van den Bosch (2016a,b), subhalo
disruption is prevalent in the Bolshoi simulation: of all subhaloes
accreted at $\zacc = 1$ (2), with a mass $\macc > 10^{-4} M_0$ (here
$M_0$ is the present-day mass of the host halo), only about 35 (10)
percent survive to the present day. And when only considering the more
massive subhaloes with $\macc/M_0 > 0.01$ ($> 0.1$) the surviving
fractions drop to roughly 20 (5) percent for an accretion redshift of
$\zacc=1$. It is interesting to compare these survival fractions with
predictions based on a disruption rate of 0.13 Gyr$^{-1}$. Using that
the look-back time to $z=1$ for the Bolshoi cosmology is 7.8 Gyr, the
expected survival fraction for subhaloes accreted at $\zacc=1$ is
equal to $0.87^{7.8} = 0.34$, in excellent agreement with the survival
fraction quoted above and taken from Jiang \& van den Bosch (2016a,b)


Based on these numbers, it appears that subhalo disruption is an
extremely important dynamical process. However, it is prudent to be
wary of numerical artifacts. After all, it is well known that limiting
mass and/or force resolution can artificially erase substructure. In
fact, this is the reason why it took until 1996 for numerical
simulations to start resolving substructure in dark matter haloes
(e.g., Moore \etal 1996; Tormen \etal 1997; Klypin \etal 1999). In the
Bolshoi simulation, the contribution of the W channel to the
disruption rate is almost certainly numerical, as it is clear that
limiting mass resolution affects the subhalo mass function for $m <
m_{50}$.  For the actual disruption channel, D, though, mass
resolution seems to be of little concern. As is evident from the
left-hand panels of Fig.~\ref{fig:massfunc}, the mass function of D
subhaloes clearly is {\it not} skewed towards low mass haloes; rather,
it is the channel with the mass function that is most heavily skewed
towards the massive end (with the possible exception of the accretion
channel A).
\begin{figure*}
\includegraphics[width=\hdsize]{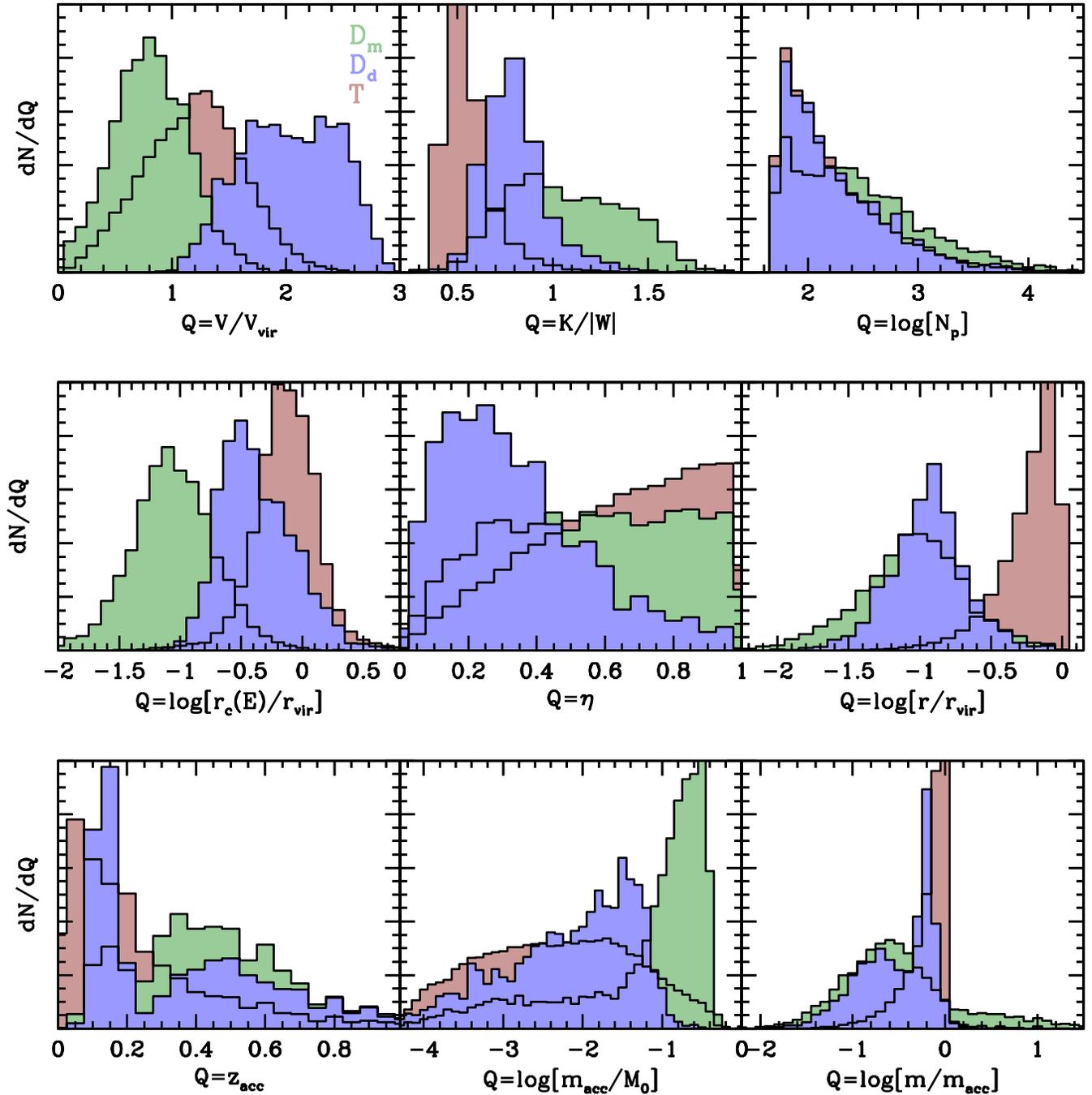}
\centering
\caption{Statistics of D subhaloes. Each panel shows the (normalized)
  distributions of various subhalo properties for \Dm subhaloes (green
  histograms), \Ds subhaloes (blue histograms) and, for comparison, T
  subhaloes (red histograms).  From top-left to bottom right, these
  subhalo properties are $V/V_{\rm vir}$, the virial ratio $K/|W|$,
  the instantaneous number of subhalo particles, $N_\rmp$, the orbital
  energy as expressed via $r_\rmc(E)/r_{\rm vir}$, the orbital
  circularity, $\eta$, the instantaneous halo-centric radius
  normalized by the host halo's virial radius, $r/r_{\rm vir}$, the
  subhalo's accretion redshift, $z_{\rm acc}$, its mass at accretion
  normalized by the present-day host halo mass, $m_{\rm acc}/M_0$, and
  the remaining mass fraction, $m/m_{\rm acc}$. Note the distinct
  demographics of \Dm and \Ds subhaloes.}
\label{fig:Dstat}
\end{figure*}

\subsubsection{Demographics of subhalo disruption}

In order to get insight as to the cause of subhalo disruption in the
Bolshoi simulation, we now examine subhaloes in the D-channel in
detail. In what follows, we exclude D subhaloes at $z < 0.012$ since,
as discussed in \S\ref{sec:DPI}, these are contaminated with phantoms,
though we have verified that including them makes no substantial
difference.

Fig.~\ref{fig:bimodal} plots the virial ratio $K/|W|$ of D-subhaloes
as a function of $V/V_{\rm vir}$, the instantaneous velocity of the
subhalo with respect to its host halo in units of the host halo's
virial velocity. The black contours indicate the 68 and 95 percentiles
of the distribution, which is clearly bimodal, suggesting that there
are two distinct populations of D-subhaloes. For comparison, the red
contours mark the 68 and 95 percentiles of the $(K/|W|,V/V_{\rm
  vir})$-distribution of T subhaloes. The dashed line roughly
separates the two categories of disrupting subhaloes, and is specified
by
\begin{equation}\label{KWcrit}
(K/|W|)_{\rm crit} = 0.9 (V/V_{\rm vir}) - 0.4
\end{equation}
In what follows we refer to subhaloes with a virial ratio $K/|W|$
larger and smaller than $(K/|W|)_{\rm crit}$ as \Dm and \Ds subhaloes,
respectively, where the subscripts refer to `merging' and
`disintegration'. Although this particular split in $(K/|W|,V/V_{\rm
  vir})$-space is somewhat arbitrary, we show below that \Dm and \Ds
subhaloes are clearly distinct. In addition, we have verified that
none of our results are sensitive to small changes in
Eq.~(\ref{KWcrit}).

As is evident from Fig.~\ref{fig:bimodal}, \Dm subhaloes have
relatively high virial ratios (median $K/|W| = 1.06$), and low
host-halo-centric velocities.  As we show below, \Dm subhaloes are
`merging' with the host halo; they loose their identity in that they
can no longer be discriminated from host halo particles in
phase-space.  Mainly due to dynamical friction, these subhaloes have
become submerged in the background that is the center of their host
halo.  Roughly 70 percent of all D-subhaloes (corresponding to a
fractional rate of 0.02 Gyr$^{-1}$) fall in this sub-category.  The
other 30 percent (corresponding to a fractional rate of 0.009
Gyr$^{-1}$) are \Ds subhaloes, which, at the moment of disruption,
have high host-halo-centric velocities, and virial ratios that are
only slightly higher than for T-subhaloes. We (provisionally)
interpret these as subhaloes that `disintegrate' due to tidal heating
and stripping.  However, as we demonstrate below, their behavior and
demographics are difficult to reconcile with these physical process,
and we suspect instead that their disruption is a numerical artifact.
\begin{figure*}
\includegraphics[width=\hdsize]{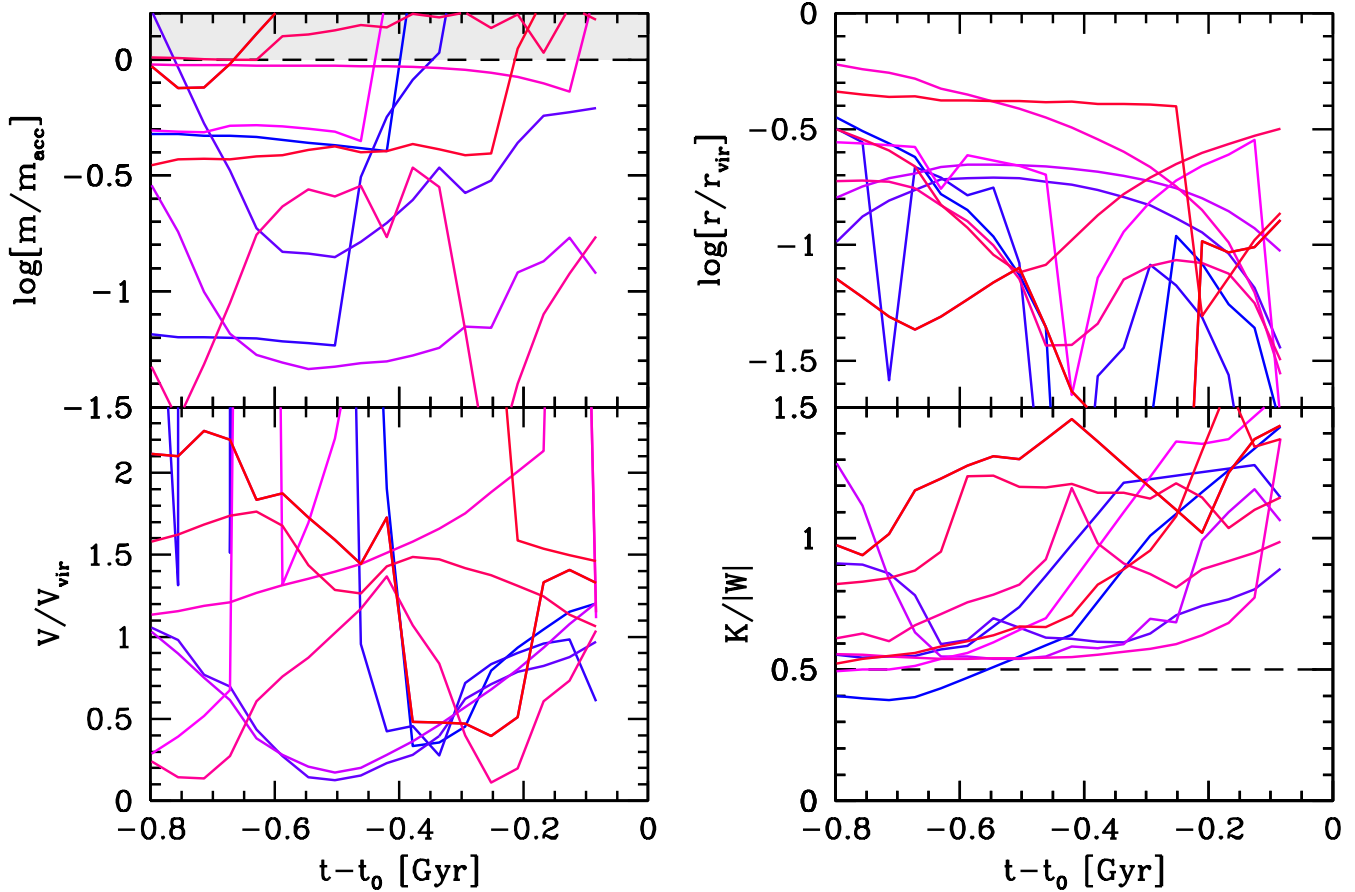}
\centering
\caption{Examples of \Dm subhaloes in Bolshoi the last 0.8 Gyr prior
  to their disruption. Colored curves show the evolution of the
  remaining mass fraction, $m/m_{\rm acc}$ (upper left-hand panel),
  the host-halo-centric radius, $r/r_{\rm vir}$ (upper right-hand
  panel), the host-halo-centric velocity normalized by the virial
  velocity of the host halo, $V/V_{\rm vir}$ (lower left-hand panel),
  and the subhalo's virial ratio, $K/|W|$ (lower right-hand panel),
  for ten randomly selected \Dm subhaloes with more than 5000
  particles at the moment of disruption. See text for detailed
  discussion.}
\label{fig:example_Dm}
\end{figure*}

Fig.~\ref{fig:Dstat} compares the distributions of nine different
parameters for \Dm subhaloes (green histograms), \Ds subhaloes (blue
histograms) and T subhaloes (red histograms). The left and middle
panels of the upper row show the distributions of $V/V_{\rm vir}$ and
$K/|W|$, which are the two parameters that were used to split
D-subhaloes in its two categories. Hence, it is not surprising that
\Dm and \Ds subhaloes appear distinct with respect to these two
parameters. However, most of the other parameters also reveal clear
differences. For example, whereas the distribution of the number of
particles, $N_\rmp$, (upper left-hand panel) of \Ds subhaloes is
virtually indistinguishable from that of T subhaloes, that of \Dm
subhaloes is offset to significantly larger values; i.e., \Dm
subhaloes are more massive than an average subhalo.  This is also
evident from the $m_{\rm acc}/M_0$ distributions, shown in the middle
panel of the lower row; 68 percent of all \Dm subhaloes have $m_{\rm
  acc}/M_0 > 0.05$, which puts them in the regime of short dynamical
friction time scales.  In terms of their orbits, \Dm subhaloes are on
orbits that are far more bound than either their \Ds counterparts, or
the `regular' T subhaloes (left-hand panel of middle row); the median
$r_\rmc(E)$ for \Dm subhaloes is only 8.6 percent of the host halo's
virial radius, compared to 36.5 percent and 74.3 percent for \Ds and T
subhaloes, respectively. In fact, the $r_\rmc(E)/r_{\rm vir}$
distribution of \Dm subhaloes is so far offset from that of T or A
subhaloes that it is clear that, on average, these subhaloes must have
lost most of their orbital energy due to dynamical friction. In terms
of the orbital circularity, $\eta$, \Ds subhaloes are clearly on very
radial orbits (middle panel of middle row). \Dm subhaloes, on the
other hand, have a fairly uniform distribution of orbital
circularities, except for an absence of the most radial orbits. As is
evident from the right-hand panel in the middle row of
Fig.~\ref{fig:Dstat}, \Ds subhaloes typically disrupt at a
halo-centric radius that is between 3 and 30 percent of the host
halo's virial radius. The distribution of $r/r_{\rm vir}$ for \Dm
subhaloes is somewhat broader.
\begin{figure*}
\includegraphics[width=\hdsize]{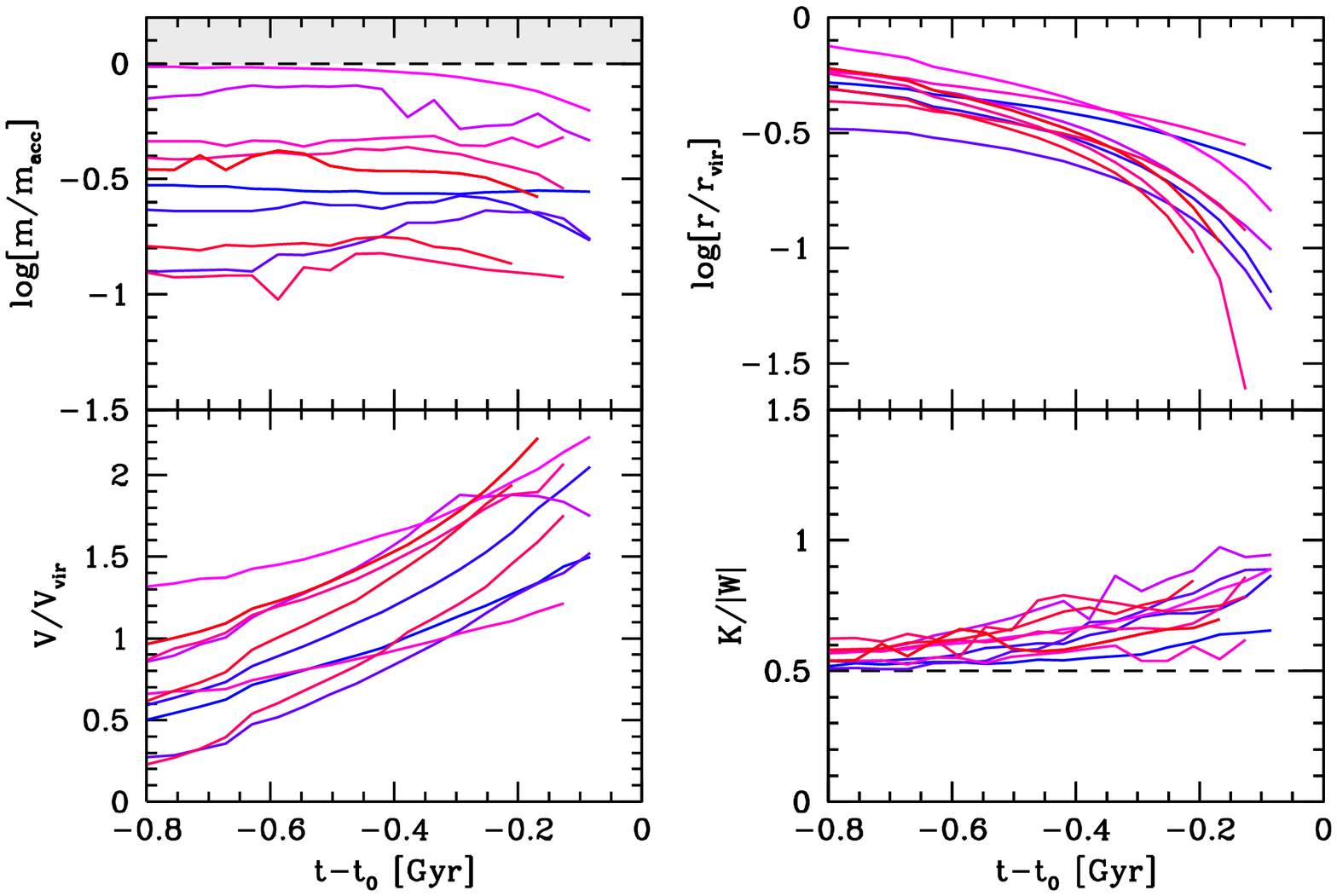}
\centering
\caption{Same as Fig.~\ref{fig:example_Dm} but for ten randomly selected
  \Ds subhaloes with more than 5000 particles at the moment of disruption.}
\label{fig:examples_Ds}
\end{figure*}

As already shown in Fig.~\ref{fig:DPI}, disrupting subhaloes have a
bimodal distribution of $z_{\rm acc}$, with two peaks around $z_{\rm
  acc} \sim 0.15$ and $0.45$. These correspond to subhaloes that are
experiencing their first and second peri-centric passages,
respectively.  The lower left-hand panel of Fig.~\ref{fig:Dstat}
reveals that the first peak is far more pronounced for \Ds subhaloes
than for \Dm subhaloes: roughly 45 percent of \Ds subhaloes disrupt
during their first orbit (i.e., have $z_{\rm acc} \leq 0.25$).  For
\Dm subhaloes this fraction is 20 percent. Finally, as evident from
the lower right-hand panel of Fig.~\ref{fig:Dstat}, a large fraction
of \Dm subhaloes (17.2 percent) have $m/m_{\rm acc} > 1$, extending
all the way to $m \sim 100 m_{\rm acc}$.  This is clearly not
physical, but rather reflects a serious problem with the proper
identification of the subhalo. As eluded to above, \Dm subhaloes are
about to merge with their host halo, which means that its phase-space
distribution `blends in' with that of its host halo.  Under such
conditions it is virtually impossible to determine for individual
particles whether they belong to the host halo or the
subhalo. Clearly, when large numbers of host-halo particles are
incorrectly associated with a subhalo (and vice versa), the properties
of these subhaloes can no longer be trusted. Hence, with the exception
of $z_{\rm acc}$ and $m_{\rm acc}/M_0$, all distributions for the \Dm
subhaloes in Fig.~\ref{fig:Dstat} are unreliable and have to be taken
with a serious grain of salt.  The $m/m_{\rm acc}$ distribution for
\Ds subhaloes is quite different from that of its \Dm
counterparts. Only $\sim 2$ percent of \Ds subhaloes have $m/m_{\rm
  acc} > 1$, and the $m/m_{\rm acc}$ distribution reveals two separate
peaks; one at $m/m_{\rm acc} \sim 0.7$ and the other at $m/m_{\rm acc}
\sim 0.2$; once again these correspond to subhaloes that disrupt at
their first and second peri-centric passage, respectively.

One of the most intriguing findings is that disrupting subhaloes are
still relatively massive. In particular, as is evident from the upper
left-hand panel of Fig.~\ref{fig:Dstat}, there are subhaloes that at
the moment of their disruption still contain more than 10,000
particles. In the case of \Dm subhaloes this is perhaps not that
surprising. After all, the subhalo needs to be massive to experience
sufficient dynamical friction. Recall, though, that the instantaneous
masses (and hence many other properties) of \Dm subhaloes are utterly
unreliable. This is also evident from Fig.~\ref{fig:example_Dm} which
plots the evolution, during the last 0.8 Gyr prior to disruption, of
various properties of ten randomly selected \Dm subhaloes that at the
moment of disruption contain more than 5,000 particles. Clearly, the
majority of these evolutionary tracks are fluctuating in a dramatic
and unphysical way, highlighting the difficulties associated with
identifying and tracking these systems.

But what are we to make of \Ds subhaloes that at the moment of
disruption still contain thousands of particles? Idealized $N$-body
simulations of individual subhaloes orbiting within an NFW host halo
suggest that whenever a dark matter subhalo disrupts, it first
experiences dramatic mass loss. The time between still having
thousands of bound particles and complete disruption is at least an
order of magnitude larger than the 42 Myr between two Bolshoi outputs.
Fig.~\ref{fig:examples_Ds} shows the same as
Fig.~\ref{fig:example_Dm}, but for ten randomly selected \Ds subhaloes
with more than 5,000 particles at the moment of disruption. These
evolutionary tracks are very different from those of the \Dm
subhaloes, once more highlighting their truly distinct nature. Prior
to their disruption, \Ds subhaloes are accelerating towards the center
of their host halo, while their virial ratio is slowly increasing.
All of this is perfectly consistent with \Ds subhaloes approaching
peri-center, where they experience a tidal shock and strong tidal
forces.  Surprisingly, however, they do not seem to experience any
significant amount of mass loss prior to their disintegration.

To summarize, there are two different `modes' of subhalo disruption in
the Bolshoi simulation (in addition to the withering described
above). \Dm subhaloes bear all the hallmarks of merging driven by
dynamical friction: they have large accretion masses (and thus short
dynamical friction time scales), are on extremely bound orbits, at
small halo-centric radii, and moving with low speed.  As a
consequence, they are difficult to identify, and most of their
properties in a halo catalog cannot be trusted. The population of \Ds
subhaloes, however, is more difficult to interpret. Their phase-space
properties are consistent with them undergoing peri-centric passage
along a highly-radial, deeply penetrating orbit. At peri-center the
subhalo is subjected to tidal heating and to a strong tidal field,
both of which are potential causes of subhalo disruption. However,
neither their masses, nor their mass histories prior to disruption,
are in line with expectations.  Especially puzzling is the large
fraction of \Ds subhaloes that disrupt at first peri-centric passage
(see also Klimentowski \etal 2010). It remains to be seen to what
extent these disruption events are physical as opposed to numerical
(due to, for example, insufficient force resolution).

\section{Summary}
\label{sec:summary}

We have presented a detailed examination of the evolution of dark
matter subhaloes in the cosmological Bolshoi simulation. We have
identified a complete set of 12 unique subhalo evolution channels, and
have analyzed the properties of subhaloes along each separate channel.
The main findings are as follows:

\begin{itemize}

\item At the moment of accretion, subhaloes are on weakly bound
  orbits, with a median $r_\rmc(E)/r_{\rm vir}$ of 1.52 and a median
  orbital circularity of $\eta = 0.49$. More than eleven percent of
  accreting subhaloes have $r_\rmc(E)/r_{\rm vir} > 3$. This is very
  different from the distributions for all subhaloes, whose median
  $r_\rmc(E)/r_{\rm vir}$ ($\eta$) is 0.74 ($0.67$), and for which
  only 1.9 percent have $r_\rmc(E)/r_{\rm vir} > 3$. Hence, there is
  very appreciable evolution in the orbital properties of subhaloes
  after accretion. This is due to three effects: (i) dynamical
  friction, which reduces $r_\rmc(E)/r_{\rm vir}$ for massive
  subhaloes, (ii) tidal disruption, which predominantly affects
  subhaloes on more radial, and more bound orbits, and (iii) growth of
  the host halo, which causes an overall decrease of $r_\rmc(E)/r_{\rm
    vir}$ of all orbits.

\item Roughly 60 percent of the present-day population of subhaloes in
  the Bolshoi simulation are on orbits whose apo-center falls outside
  of the host halo ($r_\rma > r_{\rm vir}$), giving rise to a large
  population of ejected (or backsplash) haloes (see also Lin \etal
  2003; Gill \etal 2005; Sales \etal 2007; Ludlow \etal 2009). Tidal
  forces cause a significant change in the orbital properties of these
  ejected (sub)haloes between ejection and re-accretion, mainly in the
  form of a loss of orbital angular momentum. As a consequence, a
  significant fraction of subhaloes that are ejected, and that at the
  moment of ejection are still bound to their host halo, are not
  re-accreted into the same host halo. This gives rise to `subhalo
  exchange' among neighboring host haloes. We estimate that roughly
  $0.9\pm 0.1$ percent of subhaloes changes their host halo per Gyr,
  with a weak dependence on host halo mass (see Eq.~[\ref{rateX}]).
  Hence, over a Hubble time, a significant fraction of all subhaloes
  (and thus satellite galaxies) may have changed their host halo (if
  they survive long enough).
  
\item High-speed, penetrating encounters among subhaloes are very
  common.  Hence, a significant fraction of second-order subhaloes
  (sub-subhaloes) in the Bolshoi halo catalog are actually first-order
  subhaloes undergoing an impulsive, high-speed encounter with
  another, more massive subhalo. We have estimated the rate of such
  penetrating encounters (see Eq.~[\ref{tau2enc}]) and find that
  subhaloes with a mass $m = 10^8 \Msun$ orbiting inside a Milky-Way
  size host halo have on average about one penetrating encounter with
  another, more massive subhalo per dynamical time. It remains to be
  see what impact such penetrating, high-speed encounters have on the
  structure and survivability of subhaloes and their associated
  satellite galaxies.
  
\item The mass and $\Vmax$ evolution of subhaloes in the Bolshoi
  simulation is extremely erratic, especially close to the center of
  their host halo.  Roughly 20 percent of Bolshoi subhaloes experience
  time steps in which they more than double their mass, while more
  than 40 (46) percent of all time steps result in an {\it increase}
  in mass ($\Vmax$). Most likely these fluctuations are artifacts of
  the \Rockstar halo finder that was used to analyze the Bolshoi
  simulation. \Rockstar uses 6D phase-space information to identify
  subhaloes, while the {\tt Consistent} merger tree algorithm uses
  temporal information to assure self-consistent behavior of
  subhaloes. Yet, we conclude that the instantaneous masses and
  maximum circular velocities of the resulting subhaloes are
  unreliable in that they undergo large, apparently random
  fluctuations. It remains to be seen how other subhalo finders fair
  in this respect.

\item Half of all (first-order) subhaloes at $z=0$ in the Bolshoi
  simulation were accreted onto their host haloes after $z=0.164$. And
  only 1.3 (0.02) percent of all $z=0$ subhaloes have lost more than
  90 (99) percent of their mass. The absence of subhaloes with high
  accretion redshifts and/or with small surviving mass fractions is
  evidence of efficient subhalo disruption. Indeed, among the
  (first-order) subhaloes in the Bolshoi halo catalogs we infer a
  fractional disruption rate of $\sim 0.13$ Gyr$^{-1}$. By closely
  examining the demographics of subhaloes in the Bolshoi simulation we
  have identified three different `modes' of disruption, which we
  refer to as `withering', `merging' and `disintegration'.

\item The withering mode corresponds to subhaloes that have their mass
  drop below the mass resolution limit (corresponding to 50
  particles), and they typically continue to experience mass loss
  until (shortly after withering) they disappear from the halo
  catalogs altogether (i.e., they `disrupt').  After correcting for
  the fact that a subset of these will blossom again (i.e., have their
  mass increase again above the resolution limit, mainly due to noise
  in the subhalo mass assignments), we infer a fractional withering
  rate of $\sim 0.10$ Gyr$^{-1}$, accounting for $\sim 77$ percent of
  all subhalo disruption. We emphasize that the majority of these
  disruption events are numerical, in that they occur below the mass
  resolution limit of the simulation: most of these withering
  subhaloes will survive (at least for an extensive period) if the
  simulation were to be run at higher resolution.

\item The merging mode of subhalo disruption (indicated by \Dm)
  accounts for $\sim 16$ percent of all disruption events, and
  describes subhaloes that have lost a large fraction of their orbital
  energy due to dynamical friction. At the moment of disruption they
  are on extremely bound orbits, at small halo-centric radii, and
  moving with low speed.  They `merge' with the background of the host
  halo, simply because they can no longer be identified as a
  self-bound entity, even in 6D phase-space.  As a consequence, most
  of their properties in a halo catalog (shortly before disruption)
  cannot be trusted.  Subhaloes that disrupt along this channel are
  typically very massive with more than $68$ percent having a mass at
  accretion that is larger than 5 percent of the present day host
  mass.

\item The disintegration mode of subhalo disruption, indicated by \Ds,
  accounts for the final $7$ percent of all subhalo disruption events.
  Unlike \Dm subhaloes, there is no indication that \Ds subhaloes have
  experienced appreciable dynamical friction.  Their phase-space
  properties, at the moment of disruption, are consistent with them
  undergoing peri-centric passage along a highly-radial, deeply
  penetrating orbit.  At peri-center the subhalo is subjected to tidal
  heating and a strong tidal field, both of which may potentially
  cause subhalo disruption. However, neither their masses, nor their
  mass histories prior to disruption, are in line with
  expectations. In particular, the mass function of \Ds subhaloes, at
  the moment of disruption, is indistinguishable from that of all
  subhaloes, and they experience little to no mass loss in the 0.8 Gyr
  leading up to their disruption. These aspects are difficult to
  reconcile with physical disruption due to tidal heating and/or
  stripping, and we suspect that these disruption events are also
  numerical (most likely due to inadequate force resolution).  If
  correct, this bring the fractional numerical disruption rate of
  subhaloes in the Bolshoi simulation to $0.11$ Gyr$^{-1}$, compared
  to a fractional rate of true (physical) disruption of $0.02$
  Gyr$^{-1}$.

\end{itemize}

We conclude that present-day cosmological simulations are still
subject to significant numerical over-merging. Depending on whether
the `disintegration mode' is numerical or physical, we infer that
between 77 and 84 percent of all subhalo disruption in the Bolshoi
simulation, as inferred from the publicly available \Rockstar halo
catalogs, is numerical. And this is an underestimate of the actual
artificial disruption; recall that the {\tt Consistent} merger tree
algorithm of Behroozi \etal (2013b) already attempted to correct the
halo catalogs for artificial disruption, by removing all subhaloes,
and all their progenitors, that at the moment of disruption do not
experience a tidal acceleration exceeding the limit of $|\calT| = 0.4
\kms \Myr^{-1}$ comoving Mpc$^{-1}$ (see \S\ref{sec:trees}).  Since
these systems are no longer present in the halo catalogs, they are not
included in our analysis. Consequently, the fractional disruption
rates discussed above are to be considered lower limits. We also point
out that the tidal limit of $0.4 \kms \Myr^{-1}$ comoving Mpc$^{-1}$
is fairly arbitrary, and it therefore should not come as a surprise
that a significant fraction of the remaining disruption events are
numerical as well.

The presence of significant amounts of artificial subhalo disruption
in numerical simulations has important implications. It is a serious
road-block in any attempt to use numerical simulations to interpret
clustering on small scales. In principle this may be circumvented by
allowing for `orphan' galaxies (i.e., galaxies without associated
subhaloes), something that is commonly done in semi-analytical models
for galaxy formation (e.g., Wang \etal 2006; Kitzbichler \& White
2008; Guo \etal 2011), but such a treatment is crude at best.  In
addition, over-merging in simulations is also a serious concern for
(hydro)-simulations of galaxy formation, especially if they are used
to predict properties of satellite galaxies and/or stellar haloes
(intra-cluster light). In a forthcoming paper (van den Bosch et
al. 2016, in prep.), we use a large suite of idealized numerical
simulations to examine the conditions under which subhaloes in
$N$-body simulations experience numerical and/or physical disruption,
in an attempt to shed some light on this important, outstanding issue.

\section*{Acknowledgments}

I am indebted to Peter Behroozi for all his help with the halo
catalogs of the Bolshoi simulation, and to Fangzhou Jiang, Andrew
Hearin and Duncan Campbell for valuable discussion. FvdB is supported
by the Klaus Tschira Foundation and by the US National Science
Foundation through grant AST 1516962. Part of this work was performed
at the Aspen Center for Physics, which is supported by National
Science Foundation grant PHY-1066293.



\appendix

\section{Impact of Mass Error on Subhalo Mass Function}
\label{App:convolution}

As shown in \S\ref{sec:massloss}, the mass and $\Vmax$ histories of
individual subhaloes are extremely `noisy', revealing an erratic
behavior in which subhaloes occasionally undergo large, and
unexpected,{\it increases} in either mass or maximum circular
velocity. These most likely reflect errors in the \Rockstar halo
finder. We now investigate how these errors in the instantaneous
subhalo masses impact the ability to measure a reliable subhalo mass
function.

Let $P(Q) \rmd Q$ be the distribution of $Q \equiv \log[m(t+\Delta
  t)/m(t)]$ shown in the left-hand panel of Fig.~\ref{fig:massloss}.
If we interpret time steps in which the subhalo gains mass (i.e., for
which $Q>0$) as entirely due to errors in the assigned subhalo mass,
then we may interpret the symmetric version
\begin{equation}\label{kernel}
P_{\rm sym}(Q) = \left\{ \begin{array}{ll}
    P(Q) & \mbox{if $Q \geq 0$} \\
    P(-Q) & \mbox{if $Q < 0$}
  \end{array}  \right.
\end{equation}
as a measure of the distribution of the instantaneous errors in
subhalo mass. In doing so, we interpret $Q$ as equal to $\log[1 +
  \Delta m/m]$, with $\Delta m$ the instantaneous mass error.  We can
use this to estimate the impact on the subhalo mass function, by
convolving a `true' subhalo mass function, $n_0(m) \equiv \rmd
N/\rmd(m/M_0)$, where $M_0$ is the mass of the host halo, with this
`error kernel function', according to
\begin{equation}\label{convolution}
n_{\rm conv}(m) = \int n_0(10^Q \, m) \, P(Q) \, \rmd Q\,.
\end{equation}
\begin{figure}
\includegraphics[width=\hssize]{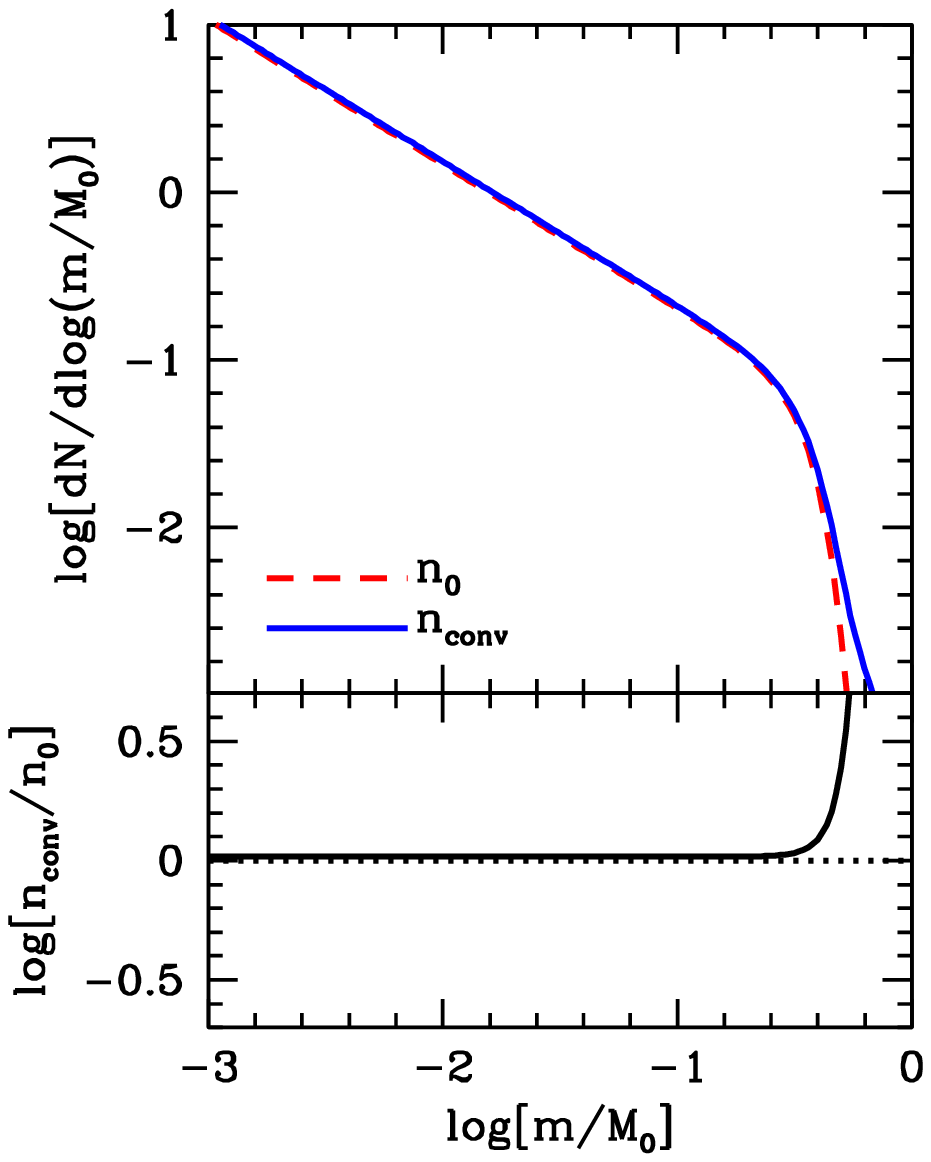}
\centering
\caption{Impact of errors on instantaneous subhalo mass on the subhalo
  mass function. The red, dashed curve in the upper panel is a typical
  subhalo mass function, $n_0 = \rmd N/\rmd(m/M_0$, described in the
  text. The blue, solid curve is the subhalo mass function one obtains
  after convolving $n_0(m)$ with the error kernel function $P(Q)$,
  which describes the probability that an instantaneous halo mass
  carries an error $\Delta m = m \, [10^{Q}-1]$. The lower panel plots
  the logarithm of the ratio $n_{\rm conv}/n(m)$.  As is evident, the
  convolved subhalo mass function is indistinguishble from the
  original one, except at the very massive end (Eddington bias).}
\label{fig:convolve}
\end{figure}

As an example, we adopt the universal functional form of the (evolved)
subhalo mass function advocated in Jiang \& van den Bosch (2016) as
representative of a typical $n_{\rm true}(m)$, and use
Eqs.~(\ref{convolution}) and~(\ref{kernel}) to compute $n_{\rm
  conv}(m)$.  The results are shown in Fig.~\ref{fig:convolve}, where
the red and blue curves correspond to $n_0(m)$ and $n_{\rm conv}(m)$,
respectively.  Clearly, the errors on $m$, as represented by $P_{\rm
  sym}(Q)$, have a completely negligible impact on the subhalo mass
function, except for the exponential tail at the massive end, where
Eddington bias (Eddington 1913) causes a systematic overestimate of
the abundance of massive subhaloes. Since that end of the mass
function is anyways uncertain due to limiting statistics (i.e., the
Poisson errors are huge), we can safely conclude that although
individual subhaloes can carry substantial errors on their
instantaneous masses, these errors have no significant impact on the
subhalo mass function. By analogy, the same applies for the subhalo
$\Vmax$ function.

\label{lastpage}

\end{document}